\documentclass[10pt,journal,final]{IEEEtran}
\IEEEoverridecommandlockouts

\usepackage{amsfonts,amssymb}
\usepackage{ifpdf}
\usepackage{cite}
\usepackage{stfloats}
\usepackage{bm}
\usepackage{color,xcolor}
\usepackage{subfigure}
\ifCLASSINFOpdf
   \usepackage[pdftex]{graphicx}
   \graphicspath{{../pdf/}{../jpeg/}}
   \DeclareGraphicsExtensions{.pdf,.jpeg,.png}
\else
   \usepackage[dvips]{graphicx}

   \graphicspath{{../eps/}}

   \DeclareGraphicsExtensions{.eps}
\fi

\usepackage[cmex10]{amsmath}

\usepackage{algorithm}
\usepackage{algorithmic}
\ifCLASSOPTIONcompsoc
\else
  \usepackage[caption=false,font=footnotesize]{subfig}
\fi

\usepackage{makecell}
\begin{document}
\title{Intelligent Reflecting Surface-Aided Joint Processing Coordinated Multipoint Transmission}
\author{
Meng Hua,
Qingqing Wu,~\IEEEmembership{Member,~IEEE,}
Derrick Wing Kwan Ng,~\IEEEmembership{Fellow,~IEEE,}
Jun Zhao,~\IEEEmembership{Member,~IEEE,}
Luxi~Yang,~\IEEEmembership{Senior Member,~IEEE}
\thanks{Manuscript received Mar    30, 2020, revised Jun   26, Oct 9, and Nov 16, and accepted Nov 27, 2020. This work was supported by the National Natural Science Foundation of China under Grants 61971128  and  U1936201, and the National Key Research and Development Program of China under Grant 2020YFB1804901.  D. W. K. Ng is supported by funding from the UNSW Digital Grid Futures Institute, UNSW, Sydney, under a cross-disciplinary fund scheme and by the Australian Research Council's Discovery Project (DP190101363). Q. Wu  is supported by the Open Research Fund of National Mobile Communications Research Laboratory, Southeast University (No. 2021D15). 
The associate editor coordinating the review of this paper and approving it for publication was Prof. Sofi\`ene Affes. (\emph{Corresponding author: Luxi Yang}.)} 
\thanks{M. Hua  and L. Yang are with the School of Information Science and Engineering, Southeast University, Nanjing 210096, China, and also with  Purple Mountain Laboratories, Nanjing 211111, China (e-mail: \{mhua, lxyang\}@seu.edu.cn).}
%\thanks{Q. Wu is with the State Key Laboratory of Internet of Things for Smart City and Department of Electrical and Computer Engineering, University of Macau, Macao, 999078 China (email: qingqingwu@um.edu.mo).}

\thanks{Q. Wu is with the State Key Laboratory of Internet of Things for Smart City, University of Macau, Macau, and also with the National Mobile Communications Research Laboratory, Southeast University, Nanjing 210096, China (email: qingqingwu@um.edu.mo). }
\thanks{D. W. K. Ng is with the School of Electrical Engineering and Telecommunications, University of New South Wales, Sydney, NSW 2052, Australia (e-mail: w.k.ng@unsw.edu.au).}
\thanks{J. Zhao is with the School of Computer Science and Engineering, Nanyang Technological University, Singapore (e-mail: junzhao@ntu.edu.sg).}
}
\maketitle
\vspace{-1cm}
\begin{abstract}
This paper investigates intelligent reflecting surface (IRS)-aided multicell wireless networks, where an IRS is deployed to assist the joint processing coordinated multipoint (JP-CoMP) transmission  from  multiple base stations (BSs) to multiple cell-edge users. By taking into account the fairness among cell-edge  users, we  aim at maximizing the minimum achievable rate of cell-edge users by jointly optimizing the transmit beamforming   at the BSs and the  phase shifts at the IRS.  As a compromise approach, we transform the non-convex max-min problem into an equivalent form based on the mean-square error  method, which facilities the design of   an efficient suboptimal iterative algorithm. In addition, we investigate two scenarios, namely the single-user system and the multiuser system.  For the former scenario, the  optimal transmit beamforming is obtained  based on the  dual subgradient  method, while  the phase shift matrix is optimized based on the Majorization-Minimization method. For the latter scenario, the transmit beamforming matrix and phase shift matrix are obtained by the second-order cone programming   and semidefinite relaxation techniques, respectively. Numerical results  demonstrate the significant performance improvement achieved by deploying an  IRS. Furthermore,  the  proposed JP-CoMP design significantly outperforms the conventional coordinated scheduling/coordinated beamforming coordinated multipoint (CS/CB-CoMP)  design in terms of max-min rate.
\end{abstract}
\section{Introduction}
To satisfy  the demands of a thousand-fold increase network capacity, several advanced technologies were proposed in the past decade, including massive multiple-input multiple-output (MIMO),  millimeter wave (mmWave) communications, and ultra-dense networks \cite{lu2014overview,larsson2014massive,swindlehurst2014millimeter,kamel2016ultra,wong2017key,2020Massivechen}. However, the energy consumption and hardware cost of the above technologies have been drastically increased  due to the substantial  power-hungry radio-frequency (RF) chains regained in   MIMO/mmWave systems  and a large number of pico/macro base stations (BSs) deployed in   ultra-dense networks \cite{zhang2017fundamental},\cite{wu2017an}. To tackle  the above issue, intelligent reflecting
surface (IRS) has been recently proposed as a promising and energy-efficient solution to improve the wireless system performance cost-effectively \cite{wu2020towards,basar2019wireless,zhao2019survey,zhang2019multiple}.

IRS is a  programmable planar surface consisting of a large number of     square metallic patch units  (e.g., low-cost printed dipoles),  each of which can be digitally controlled independently to introduce  different reflection  amplitudes, phases, polarizations, and frequency  responses on the incident signals \cite{cui2017information}, \cite{cui2014coding}. There are three main approaches to reconfigure IRS elements, such as functional materials (e.g., liquid crystal), mechanical actuation (e.g.,  mechanical rotation),  and  electronic devices (e.g., positive-intrinsic-negative (PIN) diodes, micro-electro-mechanical system (MEMS) switches, and field-effect transistors (FETs))\cite{Nayeri2018Reflectarray}. In addition,  the recently proposed  programmable metasurfaces, in which the electromagnetic  responses are manipulated by the digital-coding sequences has drawn great attentions\cite{Liaskos2018Realizing,ma2019smart,Liaskos2018anew}.  The main benefits of bringing  IRS in the future wireless networks are discussed as follows. First, each metallic patch unit is able to dynamically adjust its reflecting coefficients with the  help of a smart controller such that the desired signals and interfering signals can be  added constructively and destructively at the desired receivers, respectively\cite{huang2020Holographic,huang2019Reconfigurable}. For instance, the results  in \cite{wu2019intelligent}  showed  that for a single-user IRS-aided systems, the received signal-to-noise ratio (SNR) increases quadratically with the number of reflecting elements, $N$, at the IRS, i.e., ${\cal O}(N^2)$ which is also known as the squared power gain.  As for   multiuser systems, the multiuser  interference can be significantly suppressed by jointly optimizing the BS transmit beamforming and the IRS phase shift matrix. Second,  due to the  small structure size of a metallic patch unit, a typical IRS is capable of attaching hundreds of  such metallic patch units in practice, thereby providing a significant  beamforming gain for improving system performance. Third, since  each unit comprises passive components such as PIN diodes, varactor diodes, MEMS switches, FETs etc., to  realize  the  discrete IRS phase control or the continuous IRS phase control \cite{Zhu2013Active,Cui2016Information,Abeywickrama2020Intelligent}, the power consumption of IRS  is  thus  much lower than that of an active antenna with RF chain.  In fact, experiments conducted recently in \cite{tang2019wireless} has shown that for a large IRS consisting of $1,720$ reflecting elements, the total power consumption  is only  $0.280 \rm~W$.

Due to above appealing benefits, there have been considerable work on the development  of IRS in wireless communication systems. The existing research works about IRS include channel estimation, joint passive beamforming (i.e., IRS phase shift matrix  optimization), and BS transmit beamforming optimization. To fully reap the benefits of
the IRS in  wireless networks, acquiring  accurate channel state information (CSI) is indispensable \cite{you2019intelligent,mirza2019channel,chen2019channel}. Once the BSs have obtained the CSI, the applications of IRS to different systems have been studied to enhance their performance with different performance design objectives \cite{wu2019intelligent},\cite{wu2019beamforming,pan2019multicell,pan2019intelligent,liu2019joint,huang2018energy}.  Different from the conventional precoding adopted at the BS only, the joint optimization of the  BS transmit beamforming and the IRS phase shift matrix in IRS-aided systems is necessary  to fully unleash the potential of IRS \cite{wu2019intelligent}.  For example, an IRS-aided single-cell wireless system was studied in \cite{wu2019intelligent}, where the authors aimed at minimizing  the  transmit power at the BS by jointly optimizing the BS transmit beamforming  and the IRS passive phase matrix under the assumption  that the phase shifts at the IRS can be  continuously adjusted. It was then extended to the practical case  \cite{wu2019beamforming}, where each of the reflecting elements can  take  only finite discrete phase shift values and the results unveiled that the squared power gain can still be achieved in this case.  Besides information transmission, the applications of IRS is also appealing for substantially improving the performance of wireless power transfer systems as shown in \cite{pan2019intelligent},\cite{wu2019weighted},\cite{wu2019JointActive}. Besides, a combination of symbol-level precoding and IRS techniques for a multiuser system was studied in \cite{liu2019joint}, and a significant performance gain was obtained by the enhanced capability in mitigating.  Furthermore, it was shown in \cite{guan2020intelligent} that artificial noise can be leveraged to improve the secrecy rate in the
IRS-assisted secrecy communication, especially in presence
of multiple eavesdroppers. It is worth pointing out that the authors in \cite{huang2020ReconfigurableIntelligent} investigated  the joint design
of transmit beamforming matrix at the base station and the phase
shift matrix at the IRS  by leveraging  deep
reinforcement learning for the sum rate maximization problem.

In the past decades,  CoMP techniques have attracted great attention due to  its ability of suppressing  the intercell interference  caused  by the widely deployed pico-and macro-cells  \cite{irmer2011coordinated}.  As specified by the Third Generation Partnership Project (3GPP), there are mainly two CoMP transmission techniques: coordinated scheduling/coordinated beamforming coordinated multipoint (CS/CB-CoMP) transmission technique and joint processing coordinated multipoint (JP-CoMP) transmission technique \cite{access2010further}. For the CS/CB-CoMP transmission technique, the user data is only available at one  serving BS while the user scheduling and beamforming optimization are made with coordination among the BSs. In contrast,  for  the JP-CoMP transmission technique, the user data is  available at all BSs in the multicell network, and the BSs are capable of transmitting the same data streams to one user simultaneously \cite{irmer2011coordinated}, \cite{access2010further}. Note that
the concept of JP-CoMP is similar to that of \emph{Cell-Free Massive MIMO} with the  same objective  to achieve coherent processing across geographically distributed BSs so as to improve the system throughput \cite{Ngo2017cell},\cite{Nayebi2017precoding}. For  \emph{Cell-Free Massive MIMO} systems, the structure is relatively simple, where many  single-antenna access points (APs) simultaneously  serve a much smaller number of  single-antenna users. However, for JP-CoMP systems, the transmitters can be  equipped with multiple antennas  that  simultaneously support substantial multi-antenna users systems to improve the spectral efficiency. Furthermore, rather than deploying substantial APs in \emph{Cell-Free Massive MIMO} systems, only one BS  needs to be deployed in one cell in JP-CoMP systems, which is considerably cost-effective and energy-efficient. The question is  whether the combination of JP-CoMP technique and IRS can provide symbiotic benefits. However, this research  is still in its infancy, which motivates this work.

\begin{figure}[!t]
\centerline{\includegraphics[width=2.7in]{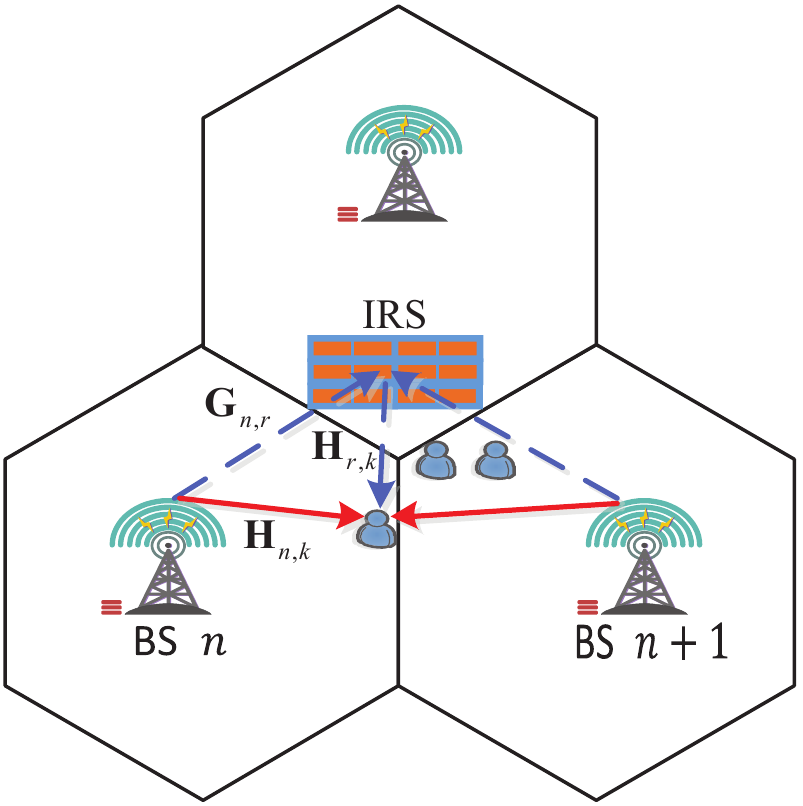}}
\caption{An IRS-aided JP-CoMP transmission multicell system.} \label{fig1}
\end{figure}

In this paper, we study an IRS-aided   JP-CoMP  downlink transmission in a  multiple-user MIMO system, where  multiple multi-antenna BSs serve multiple multi-antenna cell-edge users with the  help of an IRS. Specifically, since  cell-edge  users suffer severe propagation loss due to the long distances between them and the BSs, we deploy an IRS in the cell-edge region to  help  the BSs to serve  multiple cell-edge users. Note that an IRS can be attached to a building to  provide a high probability in establishing line-of-sight (LoS) propagation  for the BS-IRS link and IRS-user link,   as shown in Fig.~\ref{fig1}. By exploiting JP-CoMP, joint transmission can be performed among all BSs to serve the desired cell-edge users. It is observed from Fig.~\ref{fig1} that each cell-edge user  receives the superposed signals, one is  from the BSs-user link and the other is from the  BSs-IRS-user link. By carefully adapting the IRS phase shifts, multiuser interference in the system can be further  suppressed.  In addition, we compare the system performance between the considered  JP-CoMP system and small-cell systems with multicell cooperation (i.e, CS/CB-CoMP systems).  It is expected that  by fully exploiting the user data,  the intercell interference caused by the multiple BSs could be further suppressed by  JP-CoMP, thereby achieving better performance than CS/CB-CoMP.  However, it is still unknown, how much performance gain of JP-CoMP system can be achieved compared to that of CS/CB-CoMP systems with the help of IRS. In this paper, we study two different systems, namely the single user system and the multiuser system, and propose two different low-complexity suboptimal resource allocation algorithms, respectively.  The simulation results  demonstrate the superiority of our proposed  IRS-aided JP-CoMP design, and  show that our proposed IRS-aided  JP-CoMP design can achieve significantly higher performance gain compared to the existing IRS-aided  CS/CB-CoMP design. Note that our  work is different from work   \cite{pan2019multicell}, where a weighted sum rate maximization for  an IRS-aided CS/CB-CoMP transmission system was formulated. In our paper, we consider a max-min rate  maximization for an IRS-aided JP-CoMP transmission system and the results of \cite{pan2019multicell} cannot be applicable to the formulated problem in this paper, instead we
		propose two novel algorithms for single user case and multiple users case, respectively. In addition, we compare  half-duplex amplify-and-forward (AF) relay with  IRS versus the number of intelligent reflecting surface in terms of max-min rate in Section V-B. To the best of our knowledge, the JP-CoMP downlink transmission system assisted by the IRS has not been studied in the literature yet. The main contributions of this paper are summarized as follows:

\begin{itemize}
   \item We study a multicell network consisting of multiple  users, multiple BSs, and one IRS.  The BSs are connected by a central processor   for a joint data processing, and the IRS is deployed at the cell-edge region for enhancing data transmission to the  users. Taking  into account  the fairness among the users, the goal of this paper is to maximize the minimum achievable rate of the cell-edge users by jointly optimizing the  transmit beamforming matrix at the BSs and the  phase shift matrix at the IRS.
   \item  The formulated joint design problem is shown to be a non-convex optimization problem, which is difficult to solve  optimally in general. As a result, we first transform the max-min achievable rate problem into an equivalent form based on the mean-square error (MSE) method. Then, we consider two scenarios: the single-user system   and the multiuser system. For the single-user system, the BS transmit beamforming is optimally solved by the dual subgradient method when the IRS phase shift matrix is fixed, and the IRS phase shift matrix design problem is addressed  by the Majorization-Minimization (MM) method when the BS transmit beamforming is  fixed. Based on these two solutions, an efficient suboptimal iterative resource allocation algorithm   based on alternating optimization is proposed. For the multiuser system,  since the above algorithm for the single-user systems can not be applied, we transform the transmit beamforming   into a second-order cone programming  (SOCP) for  a fixed IRS phase shift matrix, which can be efficiently solved by the interior point method. In addition, for the fixed transmit beamforming matrix,  the IRS phase shift matrix is optimized  based on the semidefinite relaxation (SDR) technique. Then, an efficient iterative algorithm is   also proposed to alternately  to optimize transmit beamforming matrix and IRS phase shift matrix.

   \item   Extensive  simulations are conducted which   demonstrate that  with the assistance of an IRS, a  significant throughput gain   can be achieved compared to that without an IRS. In addition, our results also  show that the proposed IRS-aided  JP-CoMP design is superior to the IRS-aided  CS/CB-CoMP design in terms of max-min  rate.
\end{itemize}

The rest of this paper is organized as follows: Section II introduces the system model and problem formulation. In Sections III and IV, we study the IRS-aided single user and multiuser systems, respectively. Numerical results are provided in Section V, and  the paper is concluded in Section VI.

\emph{Notations:} Boldface lower-case and upper-case letter denote column vector and matrix, respectively. Transpose, conjugate, and transpose-conjugate operations are  denoted by ${\left(  \cdot  \right)^T}$, ${\left(  \cdot  \right)^*}$, and  ${\left(  \cdot  \right)^H}$, respectively. ${\mathbb C}^ {d_1\times d_2}$ stands for the set of $d_1\times d_2$ complex matrices. ${\bf I}_N$ and $\bf 0$, respectively, denote the $N\times N$ identity matrix and zero matrix.  For a square matrix $\bf Z$,  ${\rm{Tr}}\left( {\bf{Z}} \right)$, $\left| {\bf{Z}} \right|$,  ${{\bf{Z}}^{ - 1}}$, and ${\rm{rank}}\left( {\bf{Z}} \right)$ respectively, stand for  its trace, determinant, inverse, and rank, while ${\bf{Z}} \succeq {\bf{0}}$ indicates that matrix $\bf Z$ is positive semi-definite. ${\left[ {\bf{Z}} \right]_{i,i}}$ represents the $i$th diagonal element of the matrix $\bf Z$. ${\nabla _{\bf{Z}}}f\left( {\bf{Z}} \right)$ denotes  the gradient of the function $f\left( {\bf{Z}} \right)$ with respect to $\bf Z$. ${\mathop{\rm Re}\nolimits} \left(  \cdot  \right)$ denotes the real part of a complex number. $ \odot $ is a Hadamard product operator. ${\mathbb E}\left(  \cdot  \right)$ is the  expectation operator. ${\left[ x \right]^ + } = \max \left\{ {x,0} \right\}$. ${\rm arg}(\bf x)$ denotes a vector with each element being the phase of the corresponding element in $\bf x$. ${\rm diag}(\bf x) $ denotes the diagonalization operation. ${\rm{vec}}\left(  \cdot  \right)$  represents the vectorization operation. Big  ${\cal O}\left(  \cdot  \right)$ denotes the computational complexity notation.   ${\left\| {\cdot} \right\|_F}$ and ${\left\| {\cdot} \right\|_2}$ stand for the Frobenius norm and the Euclidean norm, respectively. For a complex value $e^{j\theta}$, $j$ denotes the imaginary unit. In addition, ${\bf{x}} \sim {\cal CN}\left( {{\bm{\mu }},{\bm{\Sigma }}} \right)$ denotes a circularly symmetric complex Gaussian vector with mean $\bm \mu$ and covariance matrix $\bm \Sigma$.

\section{System Model And Problem Formulation}
\subsection{System Model}
Consider an  IRS-aided   JP-CoMP downlink transmission network, which  consists of $N$ BSs, $K$ cell-edge users, and one IRS as shown in Fig.~\ref{fig1}. We assume that each  BS is equipped with $N_t>1$ transmit antennas, each cell-edge user is equipped with $N_r>1$ receiver antennas, and the IRS has $M$ reflecting elements.  Denote the sets of BSs, users, and reflecting elements as $\cal N$, $\cal K$, and $\cal M$, respectively.  We assume that the size of the considered overall area is  small so  that the delay between two paths are very small and can be neglected \cite{nigam2014coordinated,tang2018energy}. Let ${\bf{H}}_{n,k} \in {{\mathbb  C}^{N_r \times N_t}}$, ${\bf G}_{n,r} \in {{\mathbb C}^{M \times {N_t}}}$, and ${\bf H}_{r,k} \in {{\mathbb C}^{N_r \times M}}$, respectively, denote the complex equivalent baseband  channel matrix between the $k$-th user and   BS $n$, between  BS $n$ and the IRS, and  between  the  IRS and the $k$-th user,  $\forall k \in {\cal K} $, $\forall n \in {\cal N}$.

Mathematically, the transmitted signals by  BS $n$, $n \in \left\{ {1, \ldots ,N} \right\}$, is given by\footnote{In a JP-CoMP systems,   the BSs are connected to a central processing for data and information exchange among BSs so that  each user can be  served  by all the BSs simultaneously.}
\begin{align}
{{\bf{x}}_n} = \sum\limits_{k = 1}^K {{{\bf{W}}_{n,k}}{{\bf{s}}_k}}, \label{systemmodel1}
\end{align}
where ${{\bf{s}}_k} \in {{\mathbb C}^{d \times 1}}$ represents  $d$  desired  data streams for user $k$ satisfying ${{\bf{s}}_k} \sim {\cal CN}\left( {{\bf{0}},{{\bf{I}}_{d}}} \right)$, ${{\bf{W}}_{n,k}} \in {{\mathbb C}^{{N_t} \times d}}$ stands for the transmit beamforming matrix for user $k$ by BS $n$. Note that the channel acquisition techniques for the cascaded BS-IRS-user links in the IRS-aided system cannot be applied to our scenario due to the separate CSI links are required\cite{you2019intelligent,mirza2019channel,chen2019channel}.  To acquire the accurate CSI for the independent links such as BS-user links, IRS-BS links and IRS-user links, we assume that the IRS is equipped with some RF chains \cite{wu2020towards}\footnote{Note that the  IRS is passive in the sense that it is not equipped with transmit RF chains used for active transmission.}.
 As a result,  the conventional channel estimation methods for the  multicell MIMO networks can be directly  applied for  the IRS to estimate the channels of the BS-IRS and IRS-user links\footnote{ To reduce the number of receive RF
 	chains at the IRS, the sub-array technique can be applied where each sub-array consists of a cluster
 	of neighboring elements arranged vertically and/or horizontally and each cluster is equipped
 	with one receive RF chain for channel estimation. Accordingly, the reflection coefficients of all elements in each sub-array can be set to be either the same or different by applying proper interpolation over adjacent sub-arrays.  Note that at the data transmission stage, the IRS still operates in a passive way, its main energy consumption comes from the feeding circuit of the diode used to tune the phase shifts.}.
%Different from the traditional CSI estimation methods, which  the channel estimation is performed on the receiver side with substantial processing units, however, each reflecting element at the IRS is passive without powerful processing units. As a result,   the  reflecting coefficients at the IRS and transmit  pilots at the BS are jointly designed for acquiring  CSI in single-cell systems . In particular, the  CSI for IRS-aided multicell systems can be directly obtained  via activating one BS while turning off the other BSs in a take turn manner.  
As such, we assume that  the CSI for all the channel links are perfectly known by the  central processor. As shown in Fig.~\ref{fig1}, each user  receives not only the desired signals from the $N$ BSs, but also  the reflected signals by the  IRS.  Note that different from  CS/CB-CoMP multicell systems,  where each  user data is only available at one serving BS,  each user data in  JP-CoMP multicell  systems  is available at all BSs. The received signal at user $k$ is thus given by
\begin{align}
{{\bf{y}}_k} &= \;\underbrace {\sum\limits_{n = 1}^N {{{\bf{H}}_{n,k}}{{\bf{x}}_n}} }_{{\rm{direct}}{\kern 1pt} {\kern 1pt} {\kern 1pt} {\rm{links}}} + \underbrace {{{\bf{H}}_{r,k}}{\bf{\Phi }}\sum\limits_{n = 1}^N {{{\bf{G}}_{n,r}}{{\bf{x}}_n}} }_{{\rm{reflected {\kern 1pt} {\kern 1pt} {\kern 1pt} links }}} + {{\bf{n}}_k}\notag\\
&=\underbrace {\sum\limits_{n = 1}^N {\left( {{{\bf{H}}_{n,k}} + {{\bf{H}}_{r,k}}{\bf{\Phi }}{{\bf{G}}_{n,r}}} \right){{\bf{W}}_{n,k}}{{\bf{s}}_k}} }_{{\rm{desired }}{\kern 1pt} {\kern 1pt} {\rm{signals}}}+\notag\\
&~\underbrace {\sum\limits_{n = 1}^N {\sum\limits_{j \ne k}^K {\left( {{{\bf{H}}_{n,k}} + {{\bf{H}}_{r,k}}{\bf{\Phi }}{{\bf{G}}_{n,r}}} \right){{\bf{W}}_{n,j}}{{\bf{s}}_j}} } }_{{\mathop{\rm int}} {\rm{erference~signals}}} + {{\bf{n}}_k}, \label{systemmodel2}
\end{align}
where ${\bf{\Phi }} = {\rm{diag}}\left( {{a_1e^{j{\theta _1}}}, \cdots ,{a_Me^{j{\theta _M}}}} \right)$  represents  the phase shift matrix adopted at the IRS \cite{wu2019weighted,wu2020towards,pan2019intelligent,huang2019Reconfigurable,2020Intelligent}, where ${a_m} \in \left[ {0,1} \right]$ and ${\theta _m} \in \left[ {0,2\pi } \right)$ (since ${\theta _m}$ are periodic with respect to $2\pi$, thus we consider them in $\left[ {0,2\pi } \right]$ for convenience),    respectively, denote the amplitude reflection coefficient and phase shift of the $m$-th reflecting element,     ${{\bf{n}}_k} \sim {\cal CN}\left( {{\bf{0}},{\sigma ^2}{{\bf{I}}_{{N_r}}}} \right)$ is the  received noise with $\sigma^2$ denoting the  noise power at each antenna. For the sake of low implementation complexity, in this paper, each element of the IRS is designed to maximize the signal reflection, i.e., $a_m=1$, $\forall m$.

For notational simplicity, we define ${{{\bf{\bar H}}}_{n,k}} = {{\bf{H}}_{n,k}} + {{\bf{H}}_{r,k}}{\bf{\Phi }}{{\bf{G}}_{n,r}}$, ${{{\bf{\bar H}}}_k} = \left[ {{{{\bf{\bar H}}}_{1,k}}, \cdots ,{{{\bf{\bar H}}}_{N,k}}} \right]$, and ${{\bf{W}}_k} = {\left[ {{\bf{W}}_{1,k}^T, \cdots ,{\bf{W}}_{N,k}^T} \right]^T}$. Then,  we can rewrite \eqref{systemmodel2} as
\begin{align}
{{\bf{y}}_k} = {{{\bf{\bar H}}}_{k}}{{\bf{W}}_k}{{\bf{s}}_k} + {{{\bf{\bar H}}}_{k}}\sum\limits_{j \ne k}^K {{{\bf{W}}_j}{{\bf{s}}_j}}  + {{\bf{n}}_k}.
\end{align}
As such, the  achievable data rate (nat/s/Hz) of  user $k$ is given by
\begin{align}
{R_k} = \ln \left| {{\bf{I}}_{N_r} + {{{\bf{\bar H}}}_k}{{\bf{W}}_k}{\bf{W}}_k^H{\bf{\bar H}}_k^H{\bf{F}}_k^{ - 1}} \right|,
\label{systemmodel3}
\end{align}
where ${{\bf{F}}_k} = {{{\bf{\bar H}}}_k}\left( {\sum\limits_{j \ne k}^K {{{\bf{W}}_j}{\bf{W}}_j^H} } \right){\bf{\bar H}}_k^H + {\sigma ^2}{\bf{I}}_{N_r}$.
\subsection{Problem Formulation}
In this paper, to guarantee the user fairness, we aim at maximizing the minimum achievable rate of the   users by jointly optimizing the  downlink transmit beamforming and the IRS phase shift matrix, subject to transmit power constraints at the BSs\footnote{Although  having sufficient  capacity for  fronthaul  links from the BSs   to the  central processor is a challenge, some techniques such as compress-forward-estimate, estimate-compress-forward, and estimate-multiply-compress-forward are  promising approaches to address the issue of limited   fronthaul capacity \cite{kang2014joint} \cite{masoumi2020Performance}. However, to characterize the fundamental performance limits of IRS-aided JP-CoMP   systems, we assume that the capacity of fronthaul links     is sufficient for   data and CSI  exchange among BSs \cite{Ngo2017cell},\cite{Nayebi2017precoding}.}.  Accordingly, the problem can be formulated as
\begin{align}
\left( {\rm{P}} \right):&\mathop {\rm maximize }\limits_{{{\bf{W}}_{n,k}},{\bm \Phi },R} {\kern 1pt} {\kern 1pt} {\kern 1pt} {\kern 1pt} R\notag\\
&{\rm{s}}{\rm{.t}}{\rm{.}}{\kern 1pt} {\kern 1pt} {\kern 1pt} {\kern 1pt} {\kern 1pt} {\kern 1pt} \ln \left| {{\bf{I}}_{N_r} + {{{\bf{\bar H}}}_k}{{\bf{W}}_k}{\bf{W}}_k^H{\bf{\bar H}}_k^H{\bf{F}}_k^{ - 1}} \right| \ge R, {\kern 1pt} {\kern 1pt} k \in \cal K,\label{Pconst1}\\
&\qquad\sum\limits_{k = 1}^K {\left\| {{{\bf{W}}_{n,k}}} \right\|_F^2}  \le {P_{\max }}, {\kern 1pt} {\kern 1pt}n \in \cal N,\label{Pconst2}\\
&\qquad0 \le {\theta _m} \le 2\pi ,{\kern 1pt} {\kern 1pt} m \in \cal M,\label{Pconst3}
\end{align}
where $P_{\rm max}$ denotes the  maximum  BS transmit power. Although  constraint \eqref{Pconst2} is convex and \eqref{Pconst3} is linear with respect to $\theta_m$, it is  challenging to solve problem  $(\rm P)$ due to  the  coupled transmit  beamforming matrix and the phase shift matrix  in  \eqref{Pconst1}. In general, there is no efficient method to solve problem  $(\rm P)$ optimally. To facilitate the solution development, we first transform problem $(\rm P)$ into an equivalent form denoted by $(\rm P1)$ based on the mean-square error (MSE) method \cite{shi2011iteratively}. Specifically, the achievable rate in \eqref{systemmodel3} can be viewed as a data rate for a hypothetical communication system where  user $k$ estimates  the desired signal ${\bf s}_k$ with an estimator ${{\bf{U}}_k} \in {{\mathbb C}^{{N_r} \times d}}$, the estimated signal is given by
\begin{align}
{{{\bf{\hat s}}}_k} = {\bf{U}}_k^H{{\bf{y}}_k}. \label{Pconst4}
\end{align}
As such, the MSE  matrix    is given by
\begin{align}
{{\bf{E}}_k} = &{\mathbb E}\left\{ {\left( {{{{\bf{\hat s}}}_k} - {{\bf{s}}_k}} \right){{\left( {{{{\bf{\hat s}}}_k} - {{\bf{s}}_k}} \right)}^H}} \right\}\notag\\
 = &{\bf{U}}_k^H\left( {{{{\bf{\bar H}}}_k}\left( {\sum\limits_{j = 1}^K {{{\bf{W}}_j}{\bf{W}}_j^H} } \right){\bf{\bar H}}_k^H + {\sigma ^2}{\bf{I}}_{N_r}} \right){{\bf{U}}_k}\notag\\
 & -{\bf{U}}_k^H{{{\bf{\bar H}}}_k}{{\bf{W}}_k} - {\bf{W}}_k^H{\bf{\bar H}}_k^H{{\bf{U}}_k} + {\bf{I}}_{d}. \label{Pconst5}
\end{align}
By introducing additional variables ${{\bf{Q}}_k} \in {{\mathbb C}^{d \times d}}$ and ${{\bf{U}}_k} \in {{\mathbb C}^{{N_r} \times d}}$,  $\forall k$, we then  have the following theorem:

\textbf{\emph{Theorem 1:}} Problem $(\rm P)$ is equivalent\footnote{Here, ``equivalent''  means both problems share the same optimal solution.} to $(\rm P1)$, which is shown as below:
\begin{align}
\left( {\rm{P1}} \right):&\mathop {\rm maximize }\limits_{{{\bf{W}}_{n,k}},{\bm\Phi },R,{{\bf{U}}_k},{{\bf{Q}}_k}} {\kern 1pt} {\kern 1pt} {\kern 1pt} {\kern 1pt} R\notag\\
&{\rm s.t.}~\ln \left| {{{\bf{Q}}_k}} \right| - {\rm{Tr}}\left( {{{\bf{Q}}_k}{{\bf{E}}_k}} \right) + d \ge R, {\kern 1pt} {\kern 1pt} k \in \cal K,\label{P1const1}\\
& \qquad {{\bf{Q}}_k} \succeq {\bf{0}},{\kern 1pt} {\kern 1pt} k \in \cal K, \\
& \qquad \eqref{Pconst2},\eqref{Pconst3}.
\end{align}
\hspace*{\parindent}\textit{Proof}: Please refer to Appendix~A.

Although $(\rm P1)$ introduces additional variables ${{\bf{Q}}_k}$ and ${{\bf{U}}_k}$, the new problem structure facilitates the design of a computationally efficient suboptimal algorithm. In the following, we first consider a single cell-edge  user system, where the transmit beamforming matrix and phase shift matrix are obtained  based on the dual subgradient method and majorization-minimization method, respectively. Then, we consider the joint IRS phase shift and transmit  beamforming optimization problem in the multiuser system which is then handled by applying the SOCP and SDR techniques, respectively.
\section{Single Cell-edge  User System}
In this section, we consider a single cell-edge user system, namely $K=1$.   For  notational simplicity, we drop  user index $k$ in this section. Then, the problem for the single-user system can be simplified as
\begin{align}
\left( {{\rm{P2}}} \right):&\mathop {\rm maximize }\limits_{{{\bf{W}}_n},{\bm\Phi },{\bf{U}},{\bf{Q}}} {\kern 1pt} {\kern 1pt} {\kern 1pt} {\kern 1pt} \ln \left| {{{\bf{Q}}}} \right| - {\rm{Tr}}\left( {{{\bf{Q}}}{{\bf{E}}}} \right) + d\notag\\
&{\rm{s}}{\rm{.t}}{\rm{.}}{\kern 1pt} {\kern 1pt} {\kern 1pt} {\kern 1pt} {\kern 1pt} \left\| {{{\bf{W}}_n}} \right\|_F^2 \le {P_{\max }}, {\kern 1pt} {\kern 1pt} {\kern 1pt} n \in \cal N,\label{P2const1}\\
& \qquad {{\bf{Q}}} \succeq {\bf{0}},~\eqref{Pconst3}.\label{P2const1-1}
\end{align}
Although  simplified,  $(\rm P2)$ is still difficult to handle due to  the coupled optimization variables in the objective function of $(\rm P2)$. However, we observe  that both  $\bf Q$ and $\bf U$ are concave with respect to the objective function of $(\rm P2)$. In addition, variable  $\bf U$  does not exist in the constraint set and  the variable $\bf Q$  only appears  in constraint \eqref{P2const1-1}.

By applying the standard convex optimization technique, setting the first-order derivative of the objective function of $(\rm P2)$ with respective to $\bf U$  and $\bf Q$   to zero, the optimal solutions  of  $\bf U$  and $\bf Q$ can be respectively  obtained as
\begin{align}
{{\bf{U}}^{\rm opt}} = {\left( {{\bf{\bar HWW}}{{{\bf{\bar H}}}^H} + {\sigma ^2}{\bf{I}}_{N_r}} \right)^{ - 1}}{\bf{\bar HW}}, \label{P2const3}
\end{align}
and
\begin{align}
{{\bf{Q}}^{\rm opt}} = {{\bf{E}}^{ - 1}}. \label{P2const2}
\end{align}
To address the coupled transmit beamforming matrix and phase shift matrix, we first decouple $(\rm P2)$ into two sub-problems, namely transmit  beamforming optimization  with the fixed phase shift matrix and phase shift matrix optimization  with the fixed transmit beamforming matrix, and then an iterative method is proposed based on the alternating optimization \cite{shi2011iteratively}.
\subsection{Transmit    Beamforming  Matrix Optimization with Fixed Phase Shift Matrix}
We first consider the first sub-problem of $(\rm P2)$, denoted as $({\rm P2}{\rm-1})$, for optimizing the  BS transmit beamforming matrix ${\bf W}_n$ by assuming that the IRS phase shift matrix $\bm \Phi$ is  fixed. By dropping the irrelevant constant term ${\rm ln}\left| {\bf{Q}} \right| + d - {\sigma ^2}{\rm{Tr}}\left( {{\bf{Q}}{{\bf{U}}^H}{\bf{U}}} \right) - {\rm{Tr}}\left( {\bf{Q}} \right)$, the  transmit  beamforming matrix optimization problem can be simplified as
\begin{align}
&\left( {{\rm{P2}}{\rm -1}} \right):\mathop {\rm minimize }\limits_{{{\bf{W}}_n}} {\kern 1pt} {\kern 1pt} {\kern 1pt} {\kern 1pt} {\rm{Tr}}\left( {{\bf{Q}}{{\bf{U}}^H}{\bf{\bar HW}}{{\bf{W}}^H}{{{\bf{\bar H}}}^H}{\bf{U}}} \right)\notag\\
&\qquad\qquad-{\rm{Tr}}\left( {{\bf{Q}}{{\bf{U}}^H}{\bf{\bar HW}}} \right) - {\rm{Tr}}\left( {{\bf{Q}}{{\bf{W}}^H}{{{\bf{\bar H}}}^H}{\bf{U}}} \right)\notag\\
&\qquad{\rm s.t.}~\eqref{P2const1}.
\end{align}
Problem $({\rm P2}{\rm-1})$ is a standard convex optimization problem which can be solved by the convex tools such as CVX \cite{cvx}. Instead of relying on the generic solver with   high computational complexity,  we propose an efficient approach  based on the Lagrangian dual subgradient method. Note that it can be readily checked that  problem $({\rm P2}{\rm-1})$  satisfies the Slater's condition, thus, strong duality holds and its optimal solution can be obtained via solving its dual problem \cite{boyd2004convex}. In the following, we solve $({\rm P2}{\rm-1})$ by solving its dual problem.  Specifically, by introducing  dual variable $\mu_n\ge0,n\in\cal N$, corresponding to constraint \eqref{P2const1}, we have the Lagrangian function of   $({\rm P2}{\rm-1})$   given by
\begin{align}
&{\cal L}\left( {{{\bf{W}}_n},{\mu _n}} \right)= {\rm{Tr}}\left( {\left( {\sum\limits_{n = 1}^N {\sum\limits_{j = 1}^N {{{{\bf{\bar H}}}_j}{{\bf{W}}_j}{\bf{W}}_n^H{\bf{\bar H}}_n^H} } } \right){\bf{UQ}}{{\bf{U}}^H}} \right) - \notag\\
&\sum\limits_{n = 1}^N {{\rm{Tr}}\left( {{\bf{Q}}{{\bf{U}}^H}{{{\bf{\bar H}}}_n}{{\bf{W}}_n}} \right)}  - \sum\limits_{n = 1}^N {{\rm{Tr}}\left( {{\bf{QW}}_n^H{\bf{\bar H}}_n^H{\bf{U}}} \right)}  \notag\\
&+ \sum\limits_{n = 1}^N {{\mu _n}} \left( {\left\| {{{\bf{W}}_n}} \right\|_F^2 - {P_{\max }}} \right). \label{P2const4}
\end{align}
Accordingly, the dual function of $({\rm P2}{\rm-1})$  is given by
\begin{align}
g\left( {{\mu _n}} \right) = \mathop {\rm minimize }\limits_{{{\bf{W}}_n}} {\cal L}\left( {{{\bf{W}}_n},{\mu _n}} \right).
\end{align}
Setting the first-order derivative of ${\cal L}\left( {{{\bf{W}}_n},{\mu _n}} \right)$ with respect to ${{{\bf{W}}_n}}$  to zero yields
\begin{align}
{\bf{\bar H}}_n^H{\bf{UQ}}{{\bf{U}}^H}\sum\limits_{j = 1}^N {{{{\bf{\bar H}}}_j}{{\bf{W}}_j} + } {\mu _n}{{\bf{W}}_n} = {\bf{\bar H}}_n^H{\bf{UQ}}, \forall n.
\end{align}
By collecting and stacking  above $N$ equations,  the optimal transmit  beamforming matrix can be obtained as
\begin{align}
{{\bf{W}}^{\rm opt}}\left( {{\mu_n}} \right) = {\bf{\hat J}}_1^{ - 1}{{{\bf{\hat J}}}_2}, \label{P2const5}
\end{align}
where ${{{\bf{\hat J}}}_1}$ is given by
\begin{small}
\begin{align}
{{{\bf{\hat J}}}_1}=\left(                 %左括号
  \begin{array}{ccc}   %该矩阵一共3列，每一列都居中放置
    {\bf{\bar H}}_1^H{\bf{UQ}}{{\bf{U}}^H}{{{\bf{\bar H}}}_1} + {\mu _1}{\bf{I}}_{N_t} &  \cdots  & {\bf{\bar H}}_1^H{\bf{UQ}}{{\bf{U}}^H}{{{\bf{\bar H}}}_N} \\  %第一行元素
     \vdots  &  \cdots  &  \vdots \\
      {\bf{\bar H}}_N^H{\bf{UQ}}{{\bf{U}}^H}{{{\bf{\bar H}}}_1} &  \cdots  & {\bf{\bar H}}_N^H{\bf{UQ}}{{\bf{U}}^H}{{{\bf{\bar H}}}_N} + {\mu _N}{\bf{I}}_{N_t} \\
  \end{array}
\right), \label{P2const6}
\end{align}
\end{small}
and  ${{{\bf{\hat J}}}_2}$ is given by
%%%%\begin{align}
%%%%{{{\bf{\hat J}}}_2}=\left(                 %左括号
%%%%  \begin{array}{ccc}   %该矩阵一共3列，每一列都居中放置
%%%%    {\bf{\bar H}}_1^H{\bf{UQ}}\\  %第一行元素
%%%%     \vdots  \\
%%%%      {\bf{\bar H}}_N^H{\bf{UQ}} \\
%%%%  \end{array}
%%%%\right). \label{P2const7}
%%%%\end{align}
\begin{align}
{{\bf{\hat J}}_2} = {\left( {{{\left( {{\bf{H}}_1^H{\bf{UQ}}} \right)}^T}, \cdots ,{{\left( {{\bf{H}}_N^H{\bf{UQ}}} \right)}^T}} \right)^T}.\label{P2const7}
\end{align}
Next, we  address the corresponding dual problem, which is given by
\begin{align}
\left( {{\rm{P2}}}{\rm- 1}{\rm D} \right):\mathop {\rm maximize }\limits_{{\mu _n} \ge 0} g\left( {{\mu _n}} \right). \label{P2const8}
\end{align}
It can be seen that the dual problem $\left( {{\rm{P2}}}{\rm- 1}{\rm D} \right)$ has no additional constraints. In addition, with any fixed dual variable $\mu_n$, the optimal transmit beamforming matrix can be directly solved in a closed-form as in \eqref{P2const5}.
As such, we propose an efficient method, namely subgradient method, to solve the dual problem $\left( {{\rm{P2}}}{\rm- 1}{\rm D} \right)$. The update rule of parameters $\{\mu_n\}$ is given by
\begin{align}
\mu_n^{t + 1} = \left[ {\mu _n^t + {\pi _n}\left( {\left\| {{\bf{W}}_n^{\rm opt}\left( {u_n^t} \right)} \right\|_F^2 - {P_{\max }}} \right)} \right]^+, \forall n, \label{P2const9}
\end{align}
where superscript $t$ denotes the iteration index and ${\pi _n}$ represents the positive step size for updating $\mu_n$. The detailed descriptions of the dual subgradient method are summarized in Algorithm~\ref{alg1}.
\begin{algorithm}[!t]
%\algsetup{linenosize=\normalsize}
%\normalsize
\caption{Subgradient Method  for $\left( {{\rm{P2}}}{\rm-1} \right)$.}
\label{alg1}
\begin{algorithmic}[1]
\STATE  \textbf{Initialize} $\{\mu^t_n\} \ge0$, $\pi_n\ge0$, iteration index $t=0$.
\STATE  \textbf{repeat}
\STATE  \qquad Calculate the  optimal transmit beamforming matrix using  \eqref{P2const5}.
\STATE  \qquad Compute  dual variable $\{\mu_n^{t+1}\}$ using \eqref{P2const9}.
\STATE \qquad Set $t=t+1$.
\STATE \textbf{until}   the fractional increase of $\left( {{\rm{P2}}}{\rm- 1}{\rm D} \right)$ is smaller than a predefined threshold.
\STATE {\bf Output:} ${{\bf{W}}_n^{\rm opt}}$,  $\forall n \in \left\{ {1, \ldots ,N} \right\}$.
\end{algorithmic}
\end{algorithm}
\subsection{Phase Shift Matrix Optimization with Fixed Transmit Beamforming }
Next, we consider the second sub-problem of $(\rm P2)$, denoted as $({\rm P2}{\rm-2})$, for optimizing the phase shift matrix, $\bm \Phi$, by assuming that the transmit beamforming matrix, ${\bf W}_n$, is  fixed. By dropping the constant term ${\rm ln}\left| {\bf{Q}} \right| + d - {\sigma ^2}{\rm{Tr}}\left( {{\bf{Q}}{{\bf{U}}^H}{\bf{U}}} \right) - {\rm{Tr}}\left( {\bf{Q}} \right)$, the   phase shift matrix optimization  problem can be simplified as
\begin{align}
&\left( {{\rm{P2}}{\rm -2}} \right):\mathop {\rm minimize }\limits_{{\bm\Phi}} {\kern 1pt} {\kern 1pt} {\kern 1pt} {\kern 1pt} {\rm{Tr}}\left( {{\bf{Q}}{{\bf{U}}^H}{\bf{\bar HW}}{{\bf{W}}^H}{{{\bf{\bar H}}}^H}{\bf{U}}} \right)-\notag\\
&\qquad\qquad{\rm{Tr}}\left( {{\bf{Q}}{{\bf{U}}^H}{\bf{\bar HW}}} \right) - {\rm{Tr}}\left( {{\bf{Q}}{{\bf{W}}^H}{{{\bf{\bar H}}}^H}{\bf{U}}} \right)\notag\\
&\qquad\qquad{\rm s.t.}~\eqref{Pconst3}.
\end{align}
Problem $\left( {{\rm{P2}}}{\rm- 2} \right)$  is  a non-convex optimization problem due to the non-convex objective function. To address this issue, by expanding $\bf W$ and $\bf {\bar H}$, we have
\begin{align}
{\rm{Tr}}\left( {{\bf{Q}}{{\bf{U}}^H}{\bf{\bar HW}}{{\bf{W}}^H}{{{\bf{\bar H}}}^H}{\bf{U}}} \right)& = {\rm{Tr}}\left( {{{\bf{\Phi }}^H}{\bf{A\Phi \tilde E}}} \right) +{\rm{Tr}}\left( {{\bf{\Phi }}{{\bf{D}}^H}} \right)\notag\\
&+{\rm{Tr}}\left( {{{\bf{\Phi }}^H}{\bf{D}}} \right) + {c_2},\label{P-2-2cont1}
\end{align}
where ${\bf{A}} = {\bf{H}}_r^H{\bf{UQ}}{{\bf{U}}^H}{{\bf{H}}_r}$, ${\bf{\tilde E}} = \left( {\sum\limits_{n = 1}^N {{{\bf{G}}_{n,r}}{{\bf{W}}_n}} } \right){\left( {\sum\limits_{n = 1}^N {{{\bf{G}}_{n,r}}{{\bf{W}}_n}} } \right)^H}$,
${\bf{D}} = {\bf{H}}_r^H{\bf{UQ}}{{\bf{U}}^H}\left( {\sum\limits_{n = 1}^N {{{\bf{H}}_n}{{\bf{W}}_n}} } \right)$
${\left( {\sum\limits_{n = 1}^N {{{\bf{G}}_{n,r}}{{\bf{W}}_n}} } \right)^H}$, and ${c_2} = {\rm{Tr}}\left( {\left( {\sum\limits_{n = 1}^N {{{\bf{H}}_n}{{\bf{W}}_n}} } \right){{\left( {\sum\limits_{n = 1}^N {{{\bf{H}}_n}{{\bf{W}}_n}} } \right)}^H}{\bf{UQ}}{{\bf{U}}^H}} \right)$.

Similarly, we have
\begin{align}
{\rm{Tr}}\left( {{\bf{Q}}{{\bf{W}}^H}{{{\bf{\bar H}}}^H}{\bf{U}}} \right) = {\rm{Tr}}\left( {{{\bf{\Phi }}^H}{\bf{B}}} \right) + {c_1}, \label{P-2-2cont2}
\end{align}
where ${\bf{B}} = {\bf{H}}_r^H{\bf{UQ}}{\left( {\sum\limits_{n = 1}^N {{{\bf{G}}_{n,r}}{{\bf{W}}_n}} } \right)^H}$ and ${c_1} = {\rm{Tr}}\left( {{\bf{Q}}{{\left( {\sum\limits_{n = 1}^N {{{\bf{H}}_n}{{\bf{W}}_n}} } \right)}^H}{\bf{U}}} \right)$. As such, we can equivalently transform the objective function of  $\left( {{\rm{P2}}}{\rm- 2} \right)$ as (by dropping  constants $c_1$ and $c_2$)
\begin{align}
f\left( {\bf{\Phi }} \right){\rm{ = Tr}}&\left( {{{\bf{\Phi }}^H}{\bf{A\Phi \tilde E}}} \right) + {\rm{Tr}}\left( {{\bf{\Phi }}{{\left( {{\bf{D}} - {\bf{B}}} \right)}^H}} \right)+ \notag\\
&{\rm{Tr}}\left( {{{\bf{\Phi }}^H}\left( {{\bf{D}} - {\bf{B}}} \right)} \right). \label{P-2-2cont3}
\end{align}
Define  ${\bm \phi}  = \left[ {{\phi _1}, \cdots ,{\phi _M}} \right]^T$, where ${\phi _m} = {e^{j{\theta _m}}}, m\in \cal M$, and ${\bf{z}} = {\left[ {{{\left[ {{\bf{D}} - {\bf{B}}} \right]}_{1,1}}, \cdots ,{{\left[ {{\bf{D}} - {\bf{B}}} \right]}_{M,M}}} \right]^T}$.
Additionally, we have the following identities \cite{zhang2017matrix}
\begin{align}
&{\rm{Tr}}\left( {{{\bf{\Phi }}^H}{\bf{A\Phi \tilde E}}} \right) = {{\bm \phi} ^H}\left( {{\bf{A}} \odot {{{\bf{\tilde E}}}^T}} \right){\bm \phi},\notag\\
& {\rm{Tr}}\left( {{\bf{\Phi }}{{\left( {{\bf{D}} - {\bf{B}}} \right)}^H}} \right) = {{\bf{z}}^H}{\bm \phi}. \label{P-2-2cont3-1}
\end{align}
Then, we can rewrite $f\left( {\bm{\Phi }} \right)$ in \eqref{P-2-2cont3} as
\begin{align}
f\left( {\bm{\phi }} \right) = {{\bm \phi} ^H}\left( {{\bf{A}} \odot {{{\bf{\tilde E}}}^T}} \right){\bm \phi}  + {{\bf{z}}^H}{\bm \phi}  + {{\bm \phi} ^H}{\bf{z}}.\label{P-2-2cont4}
\end{align}
As  a result, problem  $\left( {{\rm{P2}}}{\rm- 2} \right)$  is equivalent to
\begin{align}
\left( {{\rm{P2}}{\rm  - 3}} \right):&\mathop {\rm minimize }\limits_{{\phi _m}} f\left( \bm \phi  \right)\notag\\
&{\rm s.t.}~\left| {{\phi _m}} \right| = 1,~m\in\cal M. \label{P-2-2cont5}
\end{align}
Problem $\left( {{\rm{P2}}}{\rm- 3} \right)$ is  non-convex due to the  unit-modulus constraints in  \eqref{P-2-2cont5}. Here, we handle $\left( {{\rm{P2}}}{\rm- 3} \right)$ based on the MM method, which guarantees at least a locally optimal  solution with a low computational complexity \cite{huang2019Reconfigurable},\cite{sun2016majorization},\cite{song2016Sequence}.  The key idea of using the  MM algorithm  lies in  constructing a sequence of convex surrogate functions. Specifically, at the $r$-th iteration, we need to construct an upper bound function of $f\left( {\bm{\phi }} \right)$, denoted as $\hat g\left( {\bm \phi |{\bm \phi ^r}} \right)$, that satisfies the following three properties \cite{sun2016majorization},\cite{song2016Sequence}: (a) $\hat g\left( {\bm \phi |{\bm \phi ^r}} \right) \ge f\left( \bm \phi  \right)$; (b) $ \hat g\left( {{\bm \phi ^r}|{\bm \phi ^r}} \right) = f\left( {{\bm \phi ^r}} \right)$;
 (c) ${\nabla _{{\bm \phi ^r}}}\hat g\left( {{\bm \phi ^r}|{\bm \phi ^r}} \right) = {\nabla _{{\bm \phi ^r}}}\left( {{\bm \phi ^r}} \right)$, where (a) denotes that  $\hat g\left( {\bm \phi |{\bm \phi ^r}} \right)$ is an  upper-bounded  function of $f\left( {\bm{\phi }} \right)$, (b) represents that  $\hat g\left( {\bm \phi |{\bm \phi ^r}} \right)$ and $f\left( {\bm{\phi }} \right)$ have the same solutions at  point  ${{\bm \phi ^r}}$, and (c) indicates $\hat g\left( {\bm \phi |{\bm \phi ^r}} \right)$ and $f\left( {\bm{\phi }} \right)$ have the same gradient at  point ${{\bm \phi ^r}}$.

Note that in \eqref{P-2-2cont4}, we can see  that $\bf A$ and ${\bf{\tilde E}}$ are semidefinite matrices, and it can be readily checked that $ \left( {{\bf{A}} \odot {{{\bf{\tilde E}}}^T}} \right)$ is also a semidefinite matrix. In the sequence, we have the following lemma:

\textbf{\emph{Lemma 1:}} Based on \cite{song2016Sequence}, at the $r$-th iteration, the surrogate function $\hat g\left( {\bm \phi |{\bm \phi ^r}} \right)$ for a quadratic function can be expressed as
\begin{align}
&\hat g\left( {\bm \phi |{\bm\phi ^r}} \right) = {\lambda _{\max }}{\bm\phi ^H}\bm\phi  - 2{\mathop{\rm Re}\nolimits} \left\{ {{\bm\phi ^H}\left( {{\lambda _{\max }}{\bf{I}}_M - {\bf{A}} \odot {{{\bf{\tilde E}}}^T}} \right){\bm\phi ^r}} \right\}\notag\\
& +{\left( {{\bm\phi ^r}} \right)^H}\left( {{\lambda _{\max }}{\bf{I}}_M - {\bf{A}} \odot {{{\bf{\tilde E}}}^T}} \right){\bm\phi ^r}  +{{\bf{z}}^H}\bm\phi  + {\bm\phi ^H}{\bf{z}}, \label{P-2-2cont6}
\end{align}
where ${\lambda _{\max }}$ is the maximum  eigenvalue of ${{\bf{A}} \odot {{{\bf{\tilde E}}}^T}}$. Therefore, at any $r$-th iteration, we solve the following problem
\begin{align}
\left( {{\rm{P2}}{\rm  - 4}} \right):&\mathop {\rm minimize}\limits_{{\bm\phi _m}} \hat g\left( {\bm\phi |{\bm\phi ^r}} \right)\notag\\
&{\rm s.t.}~\left| {{\phi _m}} \right| = 1,~m\in\cal M.
\end{align}
Since ${\bm\phi ^H}\bm\phi  = M$, at the  $r$-th iteration, we can rewrite $\hat g\left( {\bm \phi |{\bm \phi ^r}} \right)$ as
\begin{align}
\hat g\left( {\bm \phi |{\bm\phi ^r}} \right) = {\lambda _{\max }}M + 2{\mathop{\rm Re}\nolimits} \left\{ {{\bm\phi ^H}{{\bf{q}}^r}} \right\}, \label{P-2-2cont6}
\end{align}
where ${{\bf{q}}^r} = \left( {{\bf{z}} - \left( {{\lambda _{\max }}{\bf{I}}_M - {\bf{A}} \odot {{{\bf{\tilde E}}}^T}} \right){\bm\phi ^r}} \right)$. Obviously, the optimal solution $\bm \phi$ to minimize  problem $\left( {{\rm{P2}}}{\rm- 4} \right)$  is given by
\begin{align}
{{\bm \phi} ^{r,\rm opt}} = {e^{ - j\arg \left( {{{\bf{q}}^r}} \right)}}. \label{P-2-2cont7}
\end{align}
The details of the proposed MM method are summarized in Algorithm~\ref{alg2}.
\begin{algorithm}[!t]
\caption{MM Algorithm for $\left( {{\rm{P2}}}{\rm-2} \right)$.}
\label{alg2}
\begin{algorithmic}[1]
\STATE  \textbf{Initialize} ${\bm \phi ^r}$, and set iteration index $r=0$.\\
\STATE \qquad Compute the  maximum  eigenvalue of ${{\bf{A}} \odot {{{\bf{\tilde E}}}^T}}$, denoted as ${\lambda _{\max }}$.
\STATE  \textbf{repeat}
\STATE  \qquad Calculate  ${{\bf{q}}^r} = \left( {{\bf{z}} - \left( {{\lambda _{\max }}{\bf{I}}_M - {\bf{A}} \odot {{{\bf{\tilde E}}}^T}} \right){\bm \phi ^r}} \right)$.
\STATE  \qquad Obtain  the  optimal phase shift ${\bm \phi ^{r,\rm opt}}$ using  \eqref{P-2-2cont7}.
\STATE  \qquad ${\bm \phi ^{r + 1}} = {\bm\phi ^{r, \rm opt}}$.
\STATE \qquad Set $r=r+1$.
\STATE \textbf{until}  the fractional decrease of $\left( {{\rm{P2}}}{\rm- 4} \right)$ is smaller than a threshold.
\STATE {\bf Output:} $ {\bm\phi ^{\rm opt}}$.
\end{algorithmic}
\end{algorithm}
\subsection{Overall Algorithm}
\begin{algorithm}[!t]
\caption{MSE-based  Algorithm for $(\rm P2)$.}
\label{alg3}
\begin{algorithmic}[1]
\STATE  \textbf{Initialize} ${\bf W}_n$ satisfying $\left\| {{{\bf{W}}_n}} \right\|_F^2 = {P_{\max }}$.
\STATE  \textbf{repeat}
\STATE  \qquad Calculate ${{\bf{U}}^{\rm opt}}$ from \eqref{P2const3}.
\STATE  \qquad Calculate ${{\bf{Q}}^{\rm opt}}$ from \eqref{P2const2}.
\STATE  \qquad Calculate  ${{{\bf{W}}_n^{\rm opt}}}$ from Algorithm~\ref{alg1}.
\STATE  \qquad Calculate ${{\bf{\Phi}}^{\rm opt}}$ from Algorithm~\ref{alg2}.
%\STATE  \qquad ${\bm \phi ^{r + 1}} = {\bm\phi ^{r,opt}}$.
%\STATE \qquad Set $r=r+1$.
\STATE \textbf{until}  the fractional increase of the objective value of $(\rm P2)$ is less than a predefined threshold.
%\STATE {\bf Output:} $ {\bm\phi ^{opt}}$.
\end{algorithmic}
\end{algorithm}
Based on the solutions to  two sub-problems, an efficient iterative algorithm is proposed, which is summarized in  Algorithm~\ref{alg3}. To facilitate the analysis of algorithm, we define ${{\bf{U}}^{{\rm{opt,}}r}}$, ${{\bf{Q}}^{{\rm{opt,}}r}}$, ${\bf{W}}_n^{{\rm{opt,}}r}$, and ${{\bf{\Phi }}^{{\rm{opt,}}r}}$  as the solution variables in the  $r$-th iteration. 
Let $R\left( {{{\bf{U}}^{{\rm{opt,}}r}},{{\bf{Q}}^{{\rm{opt,}}r}},{\bf{W}}_n^{{\rm{opt,}}r},{{\bf{\Phi }}^{{\rm{opt,}}r}}} \right)$ and $R_{\bf{\Phi }}^{{\rm{upper}}}\left( {{{\bf{U}}^{{\rm{opt,}}r}},{{\bf{Q}}^{{\rm{opt,}}r}},{\bf{W}}_n^{{\rm{opt,}}r},{{\bf{\Phi }}^{{\rm{opt,}}r}}} \right)$ be the respective objective value of  problems $\left( {{\rm{P2}}} \right)$ and $\left( {{\rm{P2}}}{\rm- 4} \right)$. 

In the $r$-th iteration, in step 3 of Algorithm~\ref{alg3}, we  have
\begin{align}
&R\left( {{{\bf{U}}^{{\rm{opt}},r}},{{\bf{Q}}^{{\rm{opt}},r}},{\bf{W}}_n^{{\rm{opt}},r},{{\bf{\Phi }}^{{\rm{opt}},r}}} \right) \notag\\
&\le R\left( {{{\bf{U}}^{{\rm{opt}},r + 1}},{{\bf{Q}}^{{\rm{opt}},r}},{\bf{W}}_n^{{\rm{opt}},r},{{\bf{\Phi }}^{{\rm{opt}},r}}} \right), \label{singlecaseconvergence1}
\end{align}
where the inequality holds due to the fact that  $\left( {{\rm{P2}}} \right)$ is optimally solved with the fixed other variables.

Similar to  step 4 and step 5, we respectively have 
\begin{align}
&R\left( {{{\bf{U}}^{{\rm{opt}},r + 1}},{{\bf{Q}}^{{\rm{opt}},r}},{\bf{W}}_n^{{\rm{opt}},r},{{\bf{\Phi }}^{{\rm{opt}},r}}} \right) \notag\\
&\le R\left( {{{\bf{U}}^{{\rm{opt}},r + 1}},{{\bf{Q}}^{{\rm{opt}},r + 1}},{\bf{W}}_n^{{\rm{opt}},r},{{\bf{\Phi }}^{{\rm{opt}},r}}} \right),
\end{align}
and 
\begin{align}
&R\left( {{{\bf{U}}^{{\rm{opt}},r + 1}},{{\bf{Q}}^{{\rm{opt}},r + 1}},{\bf{W}}_n^{{\rm{opt}},r},{{\bf{\Phi }}^{{\rm{opt}},r}}} \right)  \notag\\
&\le R\left( {{{\bf{U}}^{{\rm{opt}},r + 1}},{{\bf{Q}}^{{\rm{opt}},r + 1}},{\bf{W}}_n^{{\rm{opt}},r + 1},{{\bf{\Phi }}^{{\rm{opt}},r}}} \right).
\end{align}
In step 6, we have 
\begin{align}
&R\left( {{{\bf{U}}^{{\rm{opt}},r + 1}},{{\bf{Q}}^{{\rm{opt}},r + 1}},{\bf{W}}_n^{{\rm{opt}},r + 1},{{\bf{\Phi }}^{{\rm{opt}},r}}} \right) \notag\\
&\overset{(a)}{=}  - R_{\bf{\Phi }}^{{\rm{upper}}}\left( {{{\bf{U}}^{{\rm{opt}},r + 1}},{{\bf{Q}}^{{\rm{opt}},r + 1}},{\bf{W}}_n^{{\rm{opt}},r + 1},{{\bf{\Phi }}^{{\rm{opt}},r}}} \right) + \Upsilon \notag\\
& \overset{(b)}{\le}   - R_{\bf{\Phi }}^{{\rm{upper}}}\left( {{{\bf{U}}^{{\rm{opt}},r + 1}},{{\bf{Q}}^{{\rm{opt}},r + 1}},{\bf{W}}_n^{{\rm{opt}},r + 1},{{\bf{\Phi }}^{{\rm{opt}},r + 1}}} \right) + \Upsilon \notag\\
&  \overset{(c)}{\le}  R\left( {{{\bf{U}}^{{\rm{opt}},r + 1}},{{\bf{Q}}^{{\rm{opt}},r + 1}},{\bf{W}}_n^{{\rm{opt}},r + 1},{{\bf{\Phi }}^{{\rm{opt}},r + 1}}} \right), \label{singlecaseconvergence4}
\end{align}
where $\Upsilon  = {\rm{ln}}\left| {\bf{Q}} \right| + d - {\sigma ^2}{\rm{Tr}}\left( {{\bf{Q}}{{\bf{U}}^H}{\bf{U}}} \right) - {\rm{Tr}}\left( {\bf{Q}} \right)$ is a constant term irrelevant to phase shift variable ${\bf{\Phi }}$ shown in Section III-B. Note that the equality $(a)$ holds since the surrogate function $\hat g\left( {\bm \phi |{\bm \phi ^r}} \right)$ in (33) is tight at the given local point ${{\bm \phi} ^r}$, which indicates that problem $\left( {{\rm{P2}}}{\rm- 3} \right)$ at ${{\bm \phi} ^r}$ has the same objective value as that of problem $\left( {{\rm{P2}}}{\rm- 4} \right)$. Besides, inequality  $(b)$ holds since problem $\left( {{\rm{P2}}}{\rm- 4} \right)$ is  optimally solved by Algorithm~\ref{alg2}. Also, $(c)$ is  due to that the objective value of $\left( {{\rm{P2}}}{\rm- 4} \right)$ is served as a  upper  bound  to   that of $\left( {{\rm{P2}}}{\rm- 3} \right)$.

Based on \eqref{singlecaseconvergence1}-\eqref{singlecaseconvergence4}, it  yields the following inequality
\begin{align}
&R\left( {{{\bf{U}}^{{\rm{opt}},r}},{{\bf{Q}}^{{\rm{opt}},r}},{\bf{W}}_n^{{\rm{opt}},r},{{\bf{\Phi }}^{{\rm{opt}},r}}} \right)\notag\\
& \le R\left( {{{\bf{U}}^{{\rm{opt}},r + 1}},{{\bf{Q}}^{{\rm{opt}},r + 1}},{\bf{W}}_n^{{\rm{opt}},r + 1},{{\bf{\Phi }}^{{\rm{opt}},r + 1}}} \right). \label{singlecaseconvergence5}
\end{align}
The inequality \eqref{singlecaseconvergence5} shows that the objective value of $\left( {{\rm{P2}}} \right)$ is non-decreasing over iterations. In addition, the maximum objective objective value of $\left( {{\rm{P2}}} \right)$ is upper bounded by  a finite value  due to the limited BS transmit power and the finite number of  IRS reflecting elements. As  such, Algorithm~\ref{alg3} is guaranteed to converge.

The complexity analysis of Algorithm~\ref{alg3} is given as below. In step 3, the complexity of computing ${{\bf{U}}^{\rm opt}}$ is ${\cal O}\left( {N_r^3} \right)$. In step 4, the complexity of computing ${{\bf{Q}}^{\rm opt}}$ is ${\cal O}\left( {d^3} \right)$. In step 5, the complexity of computing ${{\bf{W}}_n^{\rm opt}}$ is ${\cal O}{\left( {{K_\mu }N} \right)^2}$, where ${{K_\mu }}$ is  number of iterations required for updating $\mu_n$ \cite{hua2019joint},\cite{wang2015energy}. In step 6, the complexity of computing the maximum  eigenvalue, i.e., ${\lambda _{\max }}$, of ${{\bf{A}} \odot {{{\bf{\tilde E}}}^T}}$ is  ${\cal O}\left( {M^3} \right)$, and the complexity of computing ${{\bf{q}}^r}$ is ${\cal O}{\left( M^2 \right)}$, then the total complexity of  Algorithm~\ref{alg2} is ${\cal O}{\left( {{K_{\rm mm}}M} ^2+M^3\right)}$, where  ${{K_{\rm mm}}}$ is the  total number of iterations required by Algorithm~\ref{alg2} to converge. Therefore, the total complexity of Algorithm~\ref{alg3} is ${\cal O}\left( {{K_{\rm mse}}\left( {N_r^3 + {d^3} + {{\left( {{K_\mu }N} \right)}^2} + {K_{\rm mm}}{M^2} + {M^3}} \right)} \right)$, where $K_{\rm mse}$ represents the total number of iterations required by Algorithm~\ref{alg3} to converge.
\section{Multiple Cell-Edge Users System}
In this section, we consider the multiuser scenario  shown in Fig.~\ref{fig1}.  To handle problem $(\rm P1)$, it can be seen in Appendix~A, the optimal ${{\bf{Q}}_k^{\rm opt}}$  and ${\bf{U}}_k^{^{\rm opt}}$ can be directly obtained from \eqref{appendix1const4} and  \eqref{appendix1const5}, respectively. Similar to the single-user system, we decompose $(\rm P1)$ into two sub-problems, namely transmit beamforming matrix optimization  with the fixed phase shift matrix and performing phase shift matrix optimization  with the fixed transmit beamforming matrix. Note that the proposed MM method and the  dual subgradient method in the single-user system cannot be applied to the multiuser system  due to constraint \eqref{P1const1} in $(\rm P1)$. However, in the following, we resort to  SOCP technique to solve    the transmit beamforming matrix  optimization sub-problem, and the SDR technique to address the phase shift matrix optimization sub-problem.
\subsection{SOCP  for Transmit Beamforming Matrix  Optimization}
By fixing the phase shifts at the IRS, the transmit beamforming optimization problem is
\begin{align}
\left( {\rm{P1}}{\rm-1} \right):&\mathop {\rm maximize }\limits_{{{\bf{W}}_{n,k}},R} {\kern 1pt} {\kern 1pt} {\kern 1pt} {\kern 1pt} R\notag\\
&{\rm s.t.}~   \eqref{Pconst2},\eqref{P1const1}.
\end{align}
It is not difficult to observe  that $\left( {\rm{P1}}{\rm-1} \right)$ is a convex optimization problem and  can be transformed into a semidefinite program (SDP) problem. According to \cite{lobo1998applications},  the SOCP has  a much lower worst-case computational complexity than that of the SDP method by applying the interior-point method to solve problem $\left( {\rm{P1}}{\rm-1} \right)$.
We have the following theorem:

\textbf{\emph{Theorem 2:}} Problem $\left( {\rm{P1}}{\rm-1} \right)$ is equivalent to the following SOCP problem:
\begin{align}
&\left( {\rm{P1}}{\rm-2} \right):\mathop {\rm maximize }\limits_{{{\bf{W}}_{n,k}},R} {\kern 1pt} {\kern 1pt} {\kern 1pt} {\kern 1pt} R\notag\\
&{\rm s.t.}~ {\left\| {{{\bm{\eta }}_n}} \right\|_2} \le \sqrt {{P_{\max }}} ,\forall n,\\
&\qquad{\left\| {{{\bm{\omega }}_k}} \right\|_2} \le \sqrt {\ln \left| {{{\bf{Q}}_k}} \right| + d - R - {\sigma ^2}{\rm{Tr}}\left( {{{\bf{Q}}_k}{\bf{U}}_k^H{{\bf{U}}_k}} \right)} ,\forall k,
\end{align}
where ${\left\| {{{\bm{\eta }}_n}} \right\|_2}$ and ${\left\| {{{\bm{\omega }}_k}} \right\|_2}$ are, respectively, given in \eqref{appendix2const2} and \eqref{appendix2const5} in Appendix~B.

\hspace*{\parindent}\textit{Proof}: Please refer to Appendix~B.

Therefore, $\left( {\rm{P1}}{\rm-2} \right)$ is a standard SOCP problem, which can be optimally   solved by the interior point method \cite{boyd2004convex}.
\subsection{SDR Technique  for Phase Shift Matrix Optimization }
Next, by fixing the transmit beamforming matrix, the phase shift matrix optimization problem, denoted by $\left( {\rm{P1}}{\rm-3} \right)$, can be formulated as
\begin{align}
\left( {\rm{P1}}{\rm-3} \right):&\mathop {\rm maximize }\limits_{{\bm \Phi},R} {\kern 1pt} {\kern 1pt} {\kern 1pt} {\kern 1pt} R\notag\\
&{\rm s.t.}~  \eqref{Pconst3},\eqref{P1const1}.
\end{align}
Problem $\left( {\rm{P1}}{\rm-3} \right)$ is non-convex due to the non-convex  constraint  \eqref{P1const1}. To tackle this
non-convex problem, the SDR technique is applied. By using ${{{\bf{\bar H}}}_{n,k}} = {{\bf{H}}_{n,k}} + {{\bf{H}}_{r,k}}{\bf{\Phi }}{{\bf{G}}_{n,r}}$ and ${{{\bf{\bar H}}}_k} = \left[ {{{{\bf{\bar H}}}_{1,k}}, \cdots ,{{{\bf{\bar H}}}_{N,k}}} \right]$, we have
\begin{align}
&{\rm{Tr}}\left( {{{\bf{Q}}_k}{\bf{U}}_k^H{{{\bf{\bar H}}}_k}\left( {\sum\limits_{j = 1}^K {{{\bf{W}}_j}{\bf{W}}_j^H} } \right){\bf{\bar H}}_k^H{{\bf{U}}_k}} \right)\notag\\
& ={\rm{Tr}}\left( {{{\bf{\Phi }}^H}{{\bf{A}}_k}{\bf{\Phi }}{{{\bf{\check E}}}}} \right) + {\rm{Tr}}\left( {{\bf{\Phi D}}_k^H} \right) + {\rm{Tr}}\left( {{{\bf{\Phi }}^H}{{\bf{D}}_k}} \right) + {c_{k,2}},  \label{P-1-3cont1}
\end{align}
and
\begin{align}
{\rm{Tr}}\left( {{{\bf{Q}}_k}{\bf{W}}_k^H{\bf{\bar H}}_k^H{\bf{U}}_k^H} \right) = {\rm{Tr}}\left( {{{\bf{\Phi }}^H}{{\bf{B}}_k}} \right) + {c_{k,1}}, \label{P-1-3cont2}
\end{align}
where
\begin{align}
&{{\bf{A}}_k} = {\bf{H}}_{r,k}^H{{\bf{U}}_k}{{\bf{Q}}_k}{\bf{U}}_k^H{{\bf{H}}_{r,k}}, {{\bf{B}}_k} = {\bf{H}}_{r,k}^H{{\bf{U}}_k}{{\bf{Q}}_k}{\bf{L}}_{1,k}^H,\\
&\mathop {\bf{\check E}}\limits  = \sum\limits_{j = 1}^K { {{{\bf{L}}_{1,j}}{\bf{L}}_{1,j}^H}},\\
&{{\bf{D}}_k} = {\bf{H}}_{r,k}^H{{\bf{U}}_k}{{\bf{Q}}_k}{\bf{U}}_k^H\sum\limits_{j = 1}^K {{{\bf{L}}_{2,k,j}}{\bf{L}}_{1,j}^H}, \\
&{c_{k,1}} = {\rm{Tr}}\left( {{{\bf{Q}}_k}{\bf{L}}_{2,k,k}^H{{\bf{U}}_k}} \right),\\
&{c_{k,2}} = {\rm{Tr}}\left( {\sum\limits_{j = 1}^K {{{\bf{L}}_{2,k,j}}{\bf{L}}_{2,k,j}^H} {{\bf{U}}_k}{{\bf{Q}}_k}{\bf{U}}_k^H} \right),
\end{align}
where ${{\bf{L}}_{1,k}} = \sum\limits_{n = 1}^N {{{\bf{G}}_{n,r}}{{\bf{W}}_{n,k}},\forall k} $, and ${{\bf{L}}_{2,k,j}} = \sum\limits_{n = 1}^N {{{\bf{H}}_{n,k}}{{\bf{W}}_{n,j}}} ,\forall k,j$.

As a result, we can rewrite $\left( {\rm{P1}}{\rm-3} \right)$ as
\begin{align}
&\left( {\rm{P1}}{\rm-4} \right):\mathop {\max }\limits_{{\bm \Phi},R} {\kern 1pt} {\kern 1pt} {\kern 1pt} {\kern 1pt} R\notag\\
&{\rm s.t.}~ {\rm{Tr}}\left( {{{\bf{\Phi }}^H}{{\bf{A}}_k}{\bf{\Phi }}{{{\bf{\check E}}}}} \right) + {\rm{Tr}}\left( {{{\bf{\Phi }}^H}\left( {{{\bf{D}}_k} - {{\bf{B}}_k}} \right)} \right) \notag\\
&\quad+{\rm{Tr}}\left( {{\bf{\Phi }}{{\left( {{{\bf{D}}_k} - {{\bf{B}}_k}} \right)}^H}} \right) \le {\rm{cons}}{{\rm{t}}_k}-R,\forall k,\label{P-1-3cont4}\\
&\qquad\eqref{Pconst3},\notag
\end{align}
where ${\rm{cons}}{{\rm{t}}_k} = \ln \left| {{{\bf{Q}}_k}} \right| + d  + 2{\mathop{\rm Re}\nolimits} \left( {{c_{k,1}}} \right) - {c_{k,2}} - {\rm{Tr}}\left( {{{\bf{Q}}_k}\left( {{\sigma ^2}{\bf{U}}_k^H{{\bf{U}}_k} + {\bf{I}}}_d \right)} \right)$.
Similar to \eqref{P-2-2cont4},  define  ${\bm \phi}  = \left[ {{\phi _1}, \cdots ,{\phi _M}} \right]^T$, where ${\phi _m} = {e^{j{\theta _m}}}, m\in \cal M$, and ${{\bf{z}}_k} = {\left[ {{{\left[ {{{\bf{D}}_k} - {{\bf{B}}_k}} \right]}_{1,1}}, \cdots ,{{\left[ {{{\bf{D}}_k} - {{\bf{B}}_k}} \right]}_{M,M}}} \right]^T},\forall k$. Based on the identities \eqref{P-2-2cont3-1}, we thus have ${\rm{Tr}}\left( {{{\bf{\Phi }}^H}{{\bf{A}}_k}{\bm{\Phi }}{{{\bf{\check E}}}}} \right) = {\bm\phi ^H}\left( {{{\bf{A}}_k} \odot {\bf{\check E}}^T} \right)\bm\phi $, ${\rm{Tr}}\left( {{{\bf{\Phi }}^H}\left( {{{\bf{D}}_k} - {{\bf{B}}_k}} \right)} \right) = {\bm\phi ^H}{{\bf{z}}_k}$, and ${\rm{Tr}}\left( {{\bf{\Phi }}{{\left( {{{\bf{D}}_k} - {{\bf{B}}_k}} \right)}^H}} \right) = {\bf{z}}_k^H\bm\phi $. Define $\bar {\bm\phi}  = {\left[ {{{\bm\phi} ^T}{\kern 1pt} {\kern 1pt} {\kern 1pt} {\kern 1pt} {\kern 1pt} 1} \right]^T}$. Problem $\left( {\rm{P1}}{\rm-4} \right)$ is  equivalent to
\begin{align}
\left( {\rm{P1}}{\rm-5} \right):&\mathop {\rm maximize }\limits_{{\phi_m},R} {\kern 1pt} {\kern 1pt} {\kern 1pt} {\kern 1pt} R\notag\\
&{\rm s.t.}~{{\bar {\bm\phi} }^H}{{\bm{\Psi }}_k}\bar {\bm\phi}  \le {\rm{cons}}{{\rm{t}}_k} - R,\forall k,\\
&\qquad\left| {{\phi _m}} \right| = 1,\forall m,
\end{align}
where ${{\bm{\Psi }}_k}=\left(                 %左括号
  \begin{array}{ccc}   %该矩阵一共3列，每一列都居中放置
    {{{\bf{A}}_k} \odot {\bf{\check E}}^T}  &{{\bf{z}}_k} \\  %第一行元素
     {\bf{z}}_k^H &  0 \\
  \end{array}
\right)$.
However, $\left( {\rm{P1}}{\rm-5} \right)$ is still non-convex. Note that  ${{\bar {\bm \phi} }^H}{{\bm{\Psi }}_k}\bar {\bm\phi}  = {\rm{Tr}}\left( {{{\bm{\Psi }}_k}\bar {\bm \phi} {{\bar {\bm \phi} }^H}} \right)$. Define new variable ${\bf \Theta } = \bar {\bm \phi} {{\bar {\bm \phi} }^H}$, which satisfies ${\bf \Theta}  \succeq \bf 0$ and ${\rm{rank}}\left( {\bf \Theta}   \right) = 1$. Since the rank-one constraint is non-convex,  we apply SDR to relax
this constraint. The resulting problem is given by
\begin{align}
\left( {\rm{P1}}{\rm-6} \right):&\mathop {\rm maximize }\limits_{{\bf \Theta},R} {\kern 1pt} {\kern 1pt} {\kern 1pt} {\kern 1pt} R\notag\\
&{\rm s.t.}~{\rm{Tr}}\left( {{{\bm{\Psi }}_k}{\bf{\Theta }}} \right)  \le {\rm{cons}}{{\rm{t}}_k} - R,\forall k,\\
&\qquad{\bm \Theta}  \succeq \bf 0,\\
&\qquad{{\bf{\Theta }}_{m,m}} = 1,\forall m,
\end{align}
which is a standard  SDP. Therefore, $\left( {\rm{P1}}{\rm-6} \right)$ can be efficiently solved by using the  interior point methods \cite{boyd2004convex}. However, due to the relaxation for $\left( {\rm{P1}}{\rm-6} \right)$, the optimal matrix ${\bf{\Theta }}^{\rm opt}$ obtained by solving  $\left( {\rm{P1}}{\rm-6} \right)$ may not be rank-one in general. Thus, if the rank of ${\bf{\Theta }}^{\rm opt}$ is one, then we can obtain the optimal ${\bar {\bm\phi} }$ by performing  singular value decomposition on ${\bf{\Theta }}^{\rm opt}$, otherwise, we need to construct a rank-one solution from the obtained ${\bf{\Theta }}^{\rm opt}$. To address this rank-one issue, we can apply   three  effective randomization techniques proposed in \cite{sidiropoulos2006transmit} to   obtain a suboptimal solution, the details are omitted here for brevity.
\subsection{Overall Algorithm}
Based on the solutions to the above  two sub-problems, an efficient iterative approach based on the  alternating  algorithm is proposed, which is summarized in  Algorithm~\ref{alg4}.
\begin{algorithm}[!t]
\caption{Alternating optimization   for $(\rm P1)$.}
\label{alg4}
\begin{algorithmic}[1]
\STATE  \textbf{Initialize} ${\bf W}^r_{n,k}$ satisfying $\left\| {{{\bf{W}}^r_{n,k}}} \right\|_F^2 = {{{P_{\max }}} \mathord{\left/
 {\vphantom {{{P_{\max }}} K}} \right.
 \kern-\nulldelimiterspace} K}$, and $r=0$.
\STATE  \textbf{repeat}
\STATE  \quad Update ${{\bf{U}}^{{\rm opt},r+1}}$ from \eqref{appendix1const5}.
\STATE  \quad Update ${{\bf{Q}}^{{\rm opt},r+1}}$ from \eqref{appendix1const4}.
\STATE  \quad Update  ${{{\bf{W}}_{n,k}^{{\rm opt},r+1}}}$ from $\left( {\rm{P1}}{\rm-2} \right)$.
\STATE  \quad Update ${{\bf{\Phi}}^{{\rm opt},r+1}}$ from $\left( {\rm{P1}}{\rm-6} \right)$.
\STATE  \quad  If the objective value of $\left( {{\rm{P1}}}{\rm- 6} \right)$ is  smaller than that of $\left( {{\rm{P1}}}{\rm-2} \right)$, we set ${{\bf{\Phi }}^{{\rm{opt}},r + 1}}{\rm{ = }}{{\bf{\Phi }}^{{\rm{opt}},r}}$.
\STATE \quad Set $r=r+1$.
\STATE \textbf{until}  the fractional increase of the objective value of $(\rm P1)$ is less than a threshold.
%\STATE {\bf Output:} $ {\bm\phi ^{opt}}$.
\end{algorithmic}
\end{algorithm}
To facilitate the analysis of algorithm, we define ${{\bf{U}}^{{\rm{opt,}}r}}$, ${{\bf{Q}}^{{\rm{opt,}}r}}$, ${\bf{W}}_{n,k}^{{\rm{opt,}}r}$, and ${{\bf{\Phi }}^{{\rm{opt,}}r}}$ as the solution variables in the  $r$-th iteration. 
Let $\bar R\left( {{{\bf{U}}^{{\rm{opt}},r}},{{\bf{Q}}^{{\rm{opt}},r}},{\bf{W}}_{n,k}^{{\rm{opt}},r},{{\bf{\Phi }}^{{\rm{opt}},r}}} \right)$  be the  objective value of  problem $\left( {{\rm{P1}}} \right)$. 
In the $r$-th iteration, from step 3 to step 5 in  Algorithm~\ref{alg4}, we  have
\begin{align}
& \bar R\left( {{{\bf{U}}^{{\rm{opt}},r}},{{\bf{Q}}^{{\rm{opt}},r}},{\bf{W}}_{n,k}^{{\rm{opt}},r},{{\bf{\Phi }}^{{\rm{opt}},r}}} \right) \notag\\
&\le \bar R\left( {{{\bf{U}}^{{\rm{opt}},r + 1}},{{\bf{Q}}^{{\rm{opt}},r}},{\bf{W}}_{n,k}^{{\rm{opt}},r},{{\bf{\Phi }}^{{\rm{opt}},r}}} \right)  \notag\\
& \le \bar R\left( {{{\bf{U}}^{{\rm{opt}},r + 1}},{{\bf{Q}}^{{\rm{opt}},r + 1}},{\bf{W}}_{n,k}^{{\rm{opt}},r},{{\bf{\Phi }}^{{\rm{opt}},r}}} \right)\notag\\
& \le \bar R\left( {{{\bf{U}}^{{\rm{opt}},r + 1}},{{\bf{Q}}^{{\rm{opt}},r + 1}},{\bf{W}}_{n,k}^{{\rm{opt}},r + 1},{{\bf{\Phi }}^{{\rm{opt}},r}}} \right). \label{multiuserconvergence1}
\end{align}
This inequity holds due to the fact at  each step, the optimization  variable is optimally solved.
Note that in step 6, we handle problem  $\left( {{\rm{P1}}}{\rm- 6} \right)$ by using SDR method with Gaussian randomization procedure to construct a rank-one solution from the obtained ${\bf{\Theta }}^{\rm opt}$, which may not result in a  monotonic improvement property of Algorithm~\ref{alg4}. To tackle this issue, one promising solution is to perform a significant number of randomization processes  and select the best solution that maximizes the objective value of $\left( {{\rm{P1}}}{\rm- 6} \right)$. Specifically, suppose that the eigenvalue decomposition of ${\bf{\Theta }}^{\rm opt}$ is ${{\bf{\Theta }}^{{\rm{opt}}}} = {\bf{U}}_{\bf{\Theta }}{\Sigma _{\bf{\Theta }}}{\bf{V}}_{\bf{\Theta }}$  and we 
choose ${ \tilde   {\bm \phi} }$ such that  ${ \tilde   {\bm \phi} }  = {\bf{U}}_{\bf{\Theta }}^{}\Sigma _{_{\bf{\Theta }}}^{1/2}{\bf{v}}$, where ${\bf{v}}$ is  a vector of zero-mean, unit-variance complex circularly symmetric uncorrelated Gaussian random variables. We then construct a feasible solution ${{ \bar   {\bm \phi} } }$ as ${ \bar   {\bm \phi} } = \exp \left( {j\arg \left( {\frac{{{ \tilde   {\bm \phi} } }}{{{{\left[ {{ \tilde   {\bm \phi} } } \right]}_{M + 1}}}}} \right)} \right)$. In our simulation, for each iteration, we perform $1000$ random realizations for ${\bf{v}}$, and then choose   the best solution  ${ \bar   {\bm \phi} }$ that maximizes the objective value of $\left( {{\rm{P1}}}{\rm- 6} \right)$.  If the objective value of $\left( {{\rm{P1}}}{\rm- 6} \right)$ is  smaller than that of $\left( {{\rm{P1}}}{\rm-2} \right)$, we set ${{\bf{\Phi }}^{{\rm{opt}},r + 1}}{\rm{ = }}{{\bf{\Phi }}^{{\rm{opt}},r}}$ as the same solution as the last iteration.\footnote{ Note that we continue to  update all the optimization  variables by following the steps in lines 3 to 8, and  Algorithm 4 stops only when the fractional increase of the objective value of $(\rm P1)$ is less than a predefined threshold.} If  the objective value of $\left( {{\rm{P1}}}{\rm- 6} \right)$ is  larger  than that of $\left( {{\rm{P1}}}{\rm-2} \right)$, we update variable ${\bf{\Phi }}$. As a result, we always have 
\begin{align}
&\bar R\left( {{{\bf{U}}^{{\rm{opt}},r + 1}},{{\bf{Q}}^{{\rm{opt}},r + 1}},{\bf{W}}_{n,k}^{{\rm{opt}},r + 1},{{\bf{\Phi }}^{{\rm{opt}},r}}} \right) \notag\\
&\le \bar R\left( {{{\bf{U}}^{{\rm{opt}},r + 1}},{{\bf{Q}}^{{\rm{opt}},r + 1}},{\bf{W}}_{n,k}^{{\rm{opt}},r + 1},{{\bf{\Phi }}^{{\rm{opt}},r + 1}}} \right).\label{multiuserconvergence2}
\end{align}
Based on \eqref{multiuserconvergence1} and \eqref{multiuserconvergence2}, it  yields the following inequality
\begin{align}
&\bar R\left( {{{\bf{U}}^{{\rm{opt}},r }},{{\bf{Q}}^{{\rm{opt}},r + 1}},{\bf{W}}_{n,k}^{{\rm{opt}},r},{{\bf{\Phi }}^{{\rm{opt}},r}}} \right) \notag\\
&\le \bar R\left( {{{\bf{U}}^{{\rm{opt}},r + 1}},{{\bf{Q}}^{{\rm{opt}},r + 1}},{\bf{W}}_{n,k}^{{\rm{opt}},r + 1},{{\bf{\Phi }}^{{\rm{opt}},r + 1}}} \right). \label{multiusercaseconvergence}
\end{align}
The inequality \eqref{multiusercaseconvergence} shows that the objective value of $\left( {{\rm{P1}}} \right)$ is non-decreasing over iterations. In addition, the maximum objective objective value of $\left( {{\rm{P1}}} \right)$ is upper bounded by  a finite value  due to the limited BS transmit power and the finite number of  IRS reflecting elements. As such, by applying the proposed  Algorithm~\ref{alg4}, the objective value is guaranteed to be non-decreasing over the iterations and terminated finally. Note that although our proposed Algorithm~\ref{alg4} cannot guarantee to obtain a local and/or optimal solution, it provides a feasible solution to the highly non-convex $\left( {{\rm{P1}}} \right)$ with much low complexity. In addition, in Section V-B, the simulation results showed a good monotonic improvement property of Algorithm 4, which demonstrate its effectiveness.

%As  such, Algorithm~\ref{alg4} is guaranteed to converge.  {\color{red}In addition, it is worth pointing out that the converged solution is the Karush-Kuhn-Tucker (KKT) point, which  indicates that at least a locally optimal is obtained by Algorithm~\ref{alg4}. The detailed proof can be directly referred to \cite[Appendix B]{zhou2020Intelligent}, which is omitted here for brevity.}

The complexity of Algorithm~\ref{alg4} is given as follows: In step 3, the complexity of computing ${{\bf{U}}_k}$ is ${\cal O}\left( {KN_r^3} \right)$. In step 4, the complexity of computing ${{\bf{Q}}_k}$ is ${\cal O}\left( {K{d^3}} \right)$. In step 5, $\left( {\rm{P1}}{\rm-2} \right)$ is a standard  SOCP  with $2{N_t}KNd + 1$ real-valued  variables. In addition, the first constraint of $\left( {\rm{P1}}{\rm-2} \right)$ has $N$ SOC constraints, each of which has $2{N_t}Kd$ dimensions. The second constraint of  $\left( {\rm{P1}}{\rm-2} \right)$ has $K$ SOC constraints, each of which has $2N{N_t}Kd$ dimensions. Therefore,  the total complexity for solving  $\left( {\rm{P1}}{\rm-2} \right)$ is ${\cal O}\left( {\sqrt {K + N} \left( {2{N_t}KNd + 1} \right)\left( {2N{N_t}Kd + 2N{N_t}{K^2}d} \right)} \right)$, where ${\sqrt {K + N} }$ is the  number of iterations  required for  reaching convergence \cite{lobo1998applications}, \cite{pan2017joint} . In step 6, there are   $K+M+2$  number of constraints and $(M+1)^2$ complex-valued variables, thus, the complexity of solving  SDP is ${\cal O}{\left( {K +M +2+{{\left( {M + 1} \right)}^2} } \right)^{3.5}}$ \cite{sidiropoulos2006transmit}. Therefore, the total complexity of
Algorithm~\ref{alg4}  is
\begin{align}
&{\cal O}\bigg( K_{\rm alt}\Big({KN_r^3 + K{d^3} + {{\left( {K + M + {{\left( {M + 1} \right)}^2} + 2} \right)}^{3.5}} }\notag\\
&+ {\sqrt {K + N} \left( {2{N_t}KNd + 1} \right)\left( {2N{N_t}Kd + 2N{N_t}{K^2}d} \right)} \Big)\bigg),
\end{align}
 where $K_{\rm alt}$  represents the total number of iterations required by Algorithm~\ref{alg4} to converge.
\section{Numerical  results}
In this section, numerical simulations are provided to evaluate the performance of the considered IRS-aided  JP-CoMP downlink transmission system.  We assume that each BS is centered at a hexagonal cell with the side length of $200\sqrt{3}~\rm m$.  The altitudes  of the BS and the IRS are assumed to be equal with $10~\rm m$. The large-scale path loss is denoted by  ${L_{\rm loss}} = {{L}_0}{\left( {{d_x \over {{d_0}}}} \right)^{ - \alpha }}$, where ${ L}_0$ denotes the channel power gain at the reference distance ${{d_0}}=1~\rm m$,  $d_x$ is the link distance, and $\alpha$ is the path loss exponent. In our simulations, we set ${{{L}}_0} =  - 30~{\rm{dB}}$ \cite{pan2019multicell}. Since  the IRS can be attached   to the buildings, we model with LoS channels for both the BS-IRS and IRS-user links. As such,  we set the path loss exponents for the BS-IRS link, IRS-user link, and BS-user link as $\alpha_{br}=2.2$, $\alpha_{ru}=2.2$, and $\alpha_{bu}=3.6$, respectively. For the small-scale fading, we assume that the BS-IRS link and IRS-user link follow Rician fading with a Rician factor of $10~\rm dB$, and the BS-user link  follows Rayleigh fading. In addition, it is assumed that arrival of angle/departure are
randomly distributed within $[0, 2\pi]$.  Other system parameters are set as follows: $d=2$, $N_r=2$, and ${\sigma ^2} =  - 80~{\rm{dBm}}$.  Unless otherwise stated, all the results are obtained by averaging $500$ channel realizations.

For  practical IRS implementation, the phase shifters only take  a finite number of  discrete values \cite{wu2019beamforming}. Let $b$  denote the number of bits to represent the resolution levels of IRS. Then, the $m$-th discrete phase shift, denoted as ${\hat \theta _m}$,  can be derived from
\begin{align}
{{\hat \theta }_m} = \mathop {{\rm{arg}}~{\rm{min}}}\limits_{\theta  \in {\cal F}} \left| {{e^{j\theta }} - {e^{j\theta _m^{{\rm{opt}}}}}} \right|,
\end{align}
where ${\cal F} = \left\{ {0,{{2\pi } \mathord{\left/
 {\vphantom {{2\pi } {{2^b}, \ldots ,{{2\pi \left( {{2^b} - 1} \right)} \mathord{\left/
 {\vphantom {{2\pi \left( {{2^b} - 1} \right)} {{2^b}}}} \right.
 \kern-\nulldelimiterspace} {{2^b}}}}}} \right.
 \kern-\nulldelimiterspace} {{2^b}, \ldots ,{{2\pi \left( {{2^b} - 1} \right)} \mathord{\left/
 {\vphantom {{2\pi \left( {{2^b} - 1} \right)} {{2^b}}}} \right.
 \kern-\nulldelimiterspace} {{2^b}}}}}} \right\}$, and ${\theta _m^{\rm opt}}$ denotes the continuous   phase shift at the $m$-th reflecting element obtained  by solving the   proposed Algorithm~\ref{alg3} for the single-user system and Algorithm~\ref{alg4} for the multiuser system.
\subsection{Single-user System}
\begin{figure}[!t]
\centering
\begin{minipage}[t]{0.4\textwidth}
	\centering
	\includegraphics[width=2.8in]{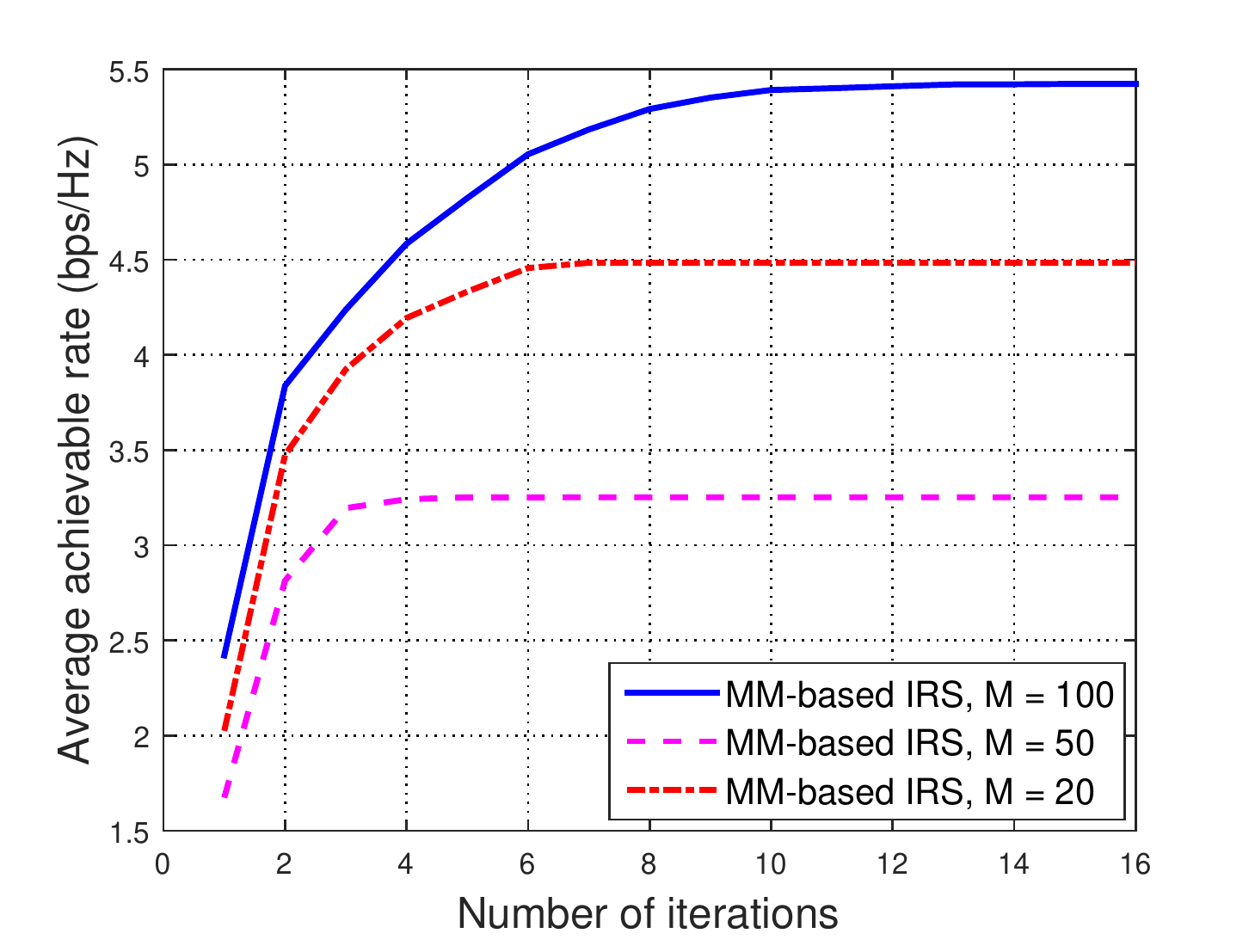}
	\caption{Convergence behaviour of  Algorithm~\ref{alg3}.}\label{fig2}
\end{minipage}
\begin{minipage}[t]{0.45\textwidth}
	\centering
	\includegraphics[width=2.8in]{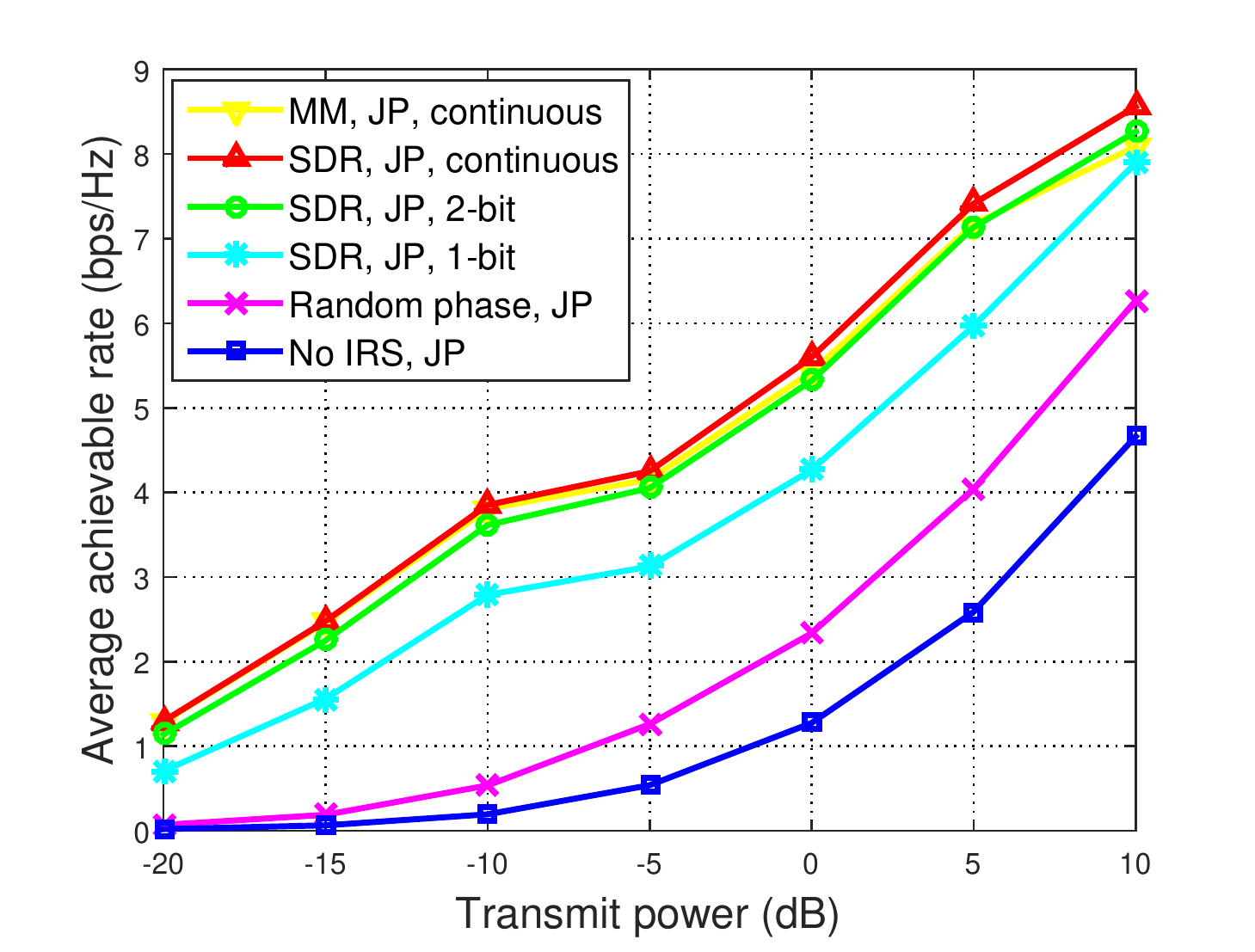}
	\caption{Average achievable rate versus  BS transmit power budget.}\label{fig3}
\end{minipage}
\end{figure}

We first consider a system  with only one cell-edge user. We assume that there are two BSs,  which are respectively located at  $(-300 ~\rm m, 0)$ and $(300~ \rm m, 0)$ in the horizontal plane.
We assume that the user is located at the middle of a line connecting two BSs, i.e., the user is located at  $(0, 0)$.
Before the performance comparison, we first show the convergence behaviour of
Algorithm~\ref{alg3} for the single-user system as shown in Fig.~\ref{fig2}. In particular,   we show the average  achievable rate (average max-min rate for the single-user systems) versus the number of iterations for the different number of IRS reflecting elements, namely $M=20$, $M=50$, and $M=100$, under   $N_t=2$ and $P_{\rm max}=1~\rm W$. It is observed that the average achievable rate obtained by the different number of reflecting elements all  increases quickly with the number of iterations. For a large number of reflecting elements, i.e.,  $M=100$, the proposed algorithm converges within   $10$ iterations. Especially, for a small number of reflecting elements, i.e.,  $M=20$, the proposed algorithm converges in about  $4$ iterations as the feasible solution set is smaller. This  demonstrates the fast convergence  of the   proposed Algorithm~\ref{alg3}.

\begin{figure}[!t]
\centering
\begin{minipage}[t]{0.45\textwidth}
\centering
\includegraphics[width=2.8in]{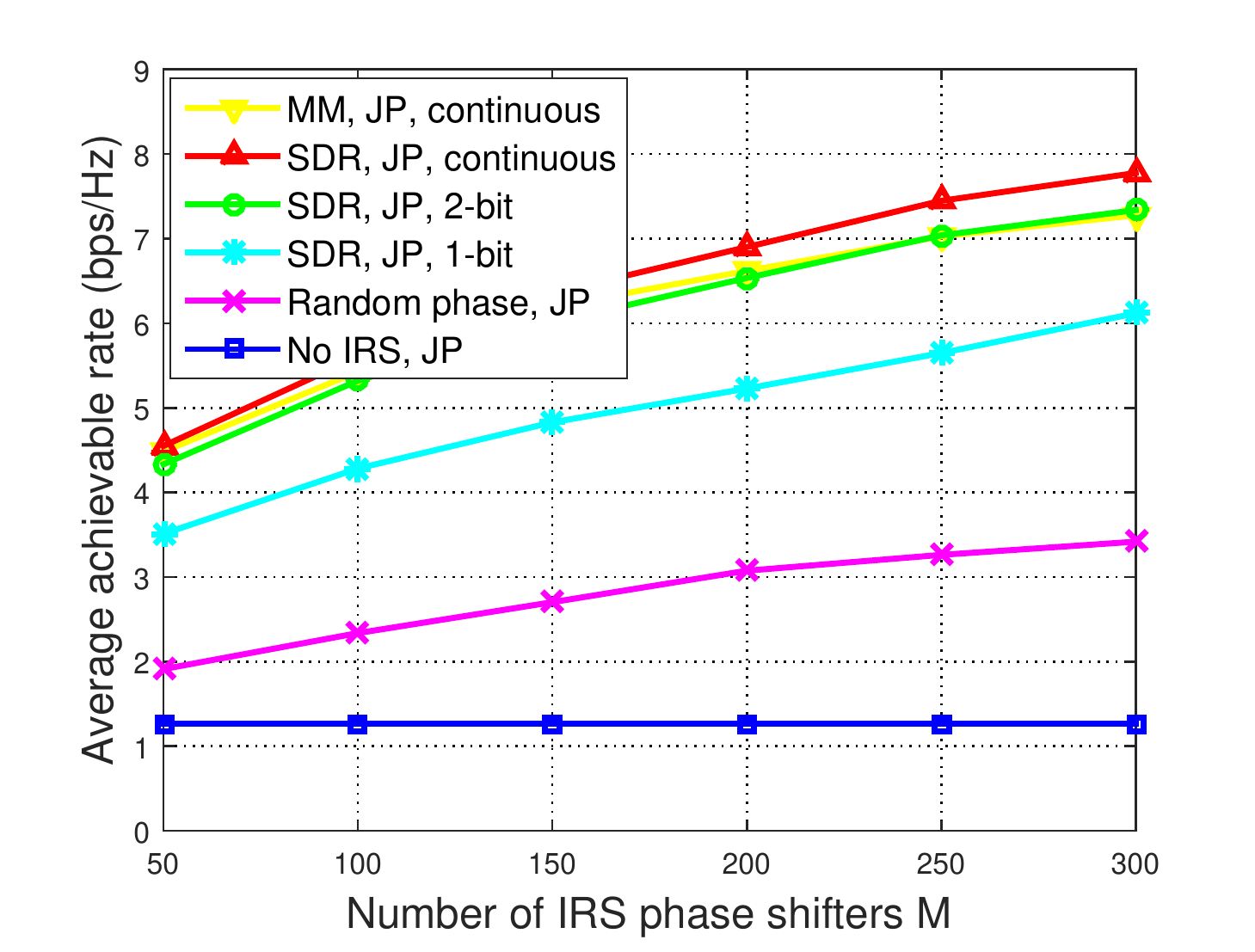}
\caption{Average achievable rate versus  the number of IRS reflecting elements under $N_t=2$.}\label{fig5} 
\end{minipage}
\begin{minipage}[t]{0.45\textwidth}
	\centering
	\includegraphics[width=2.8in]{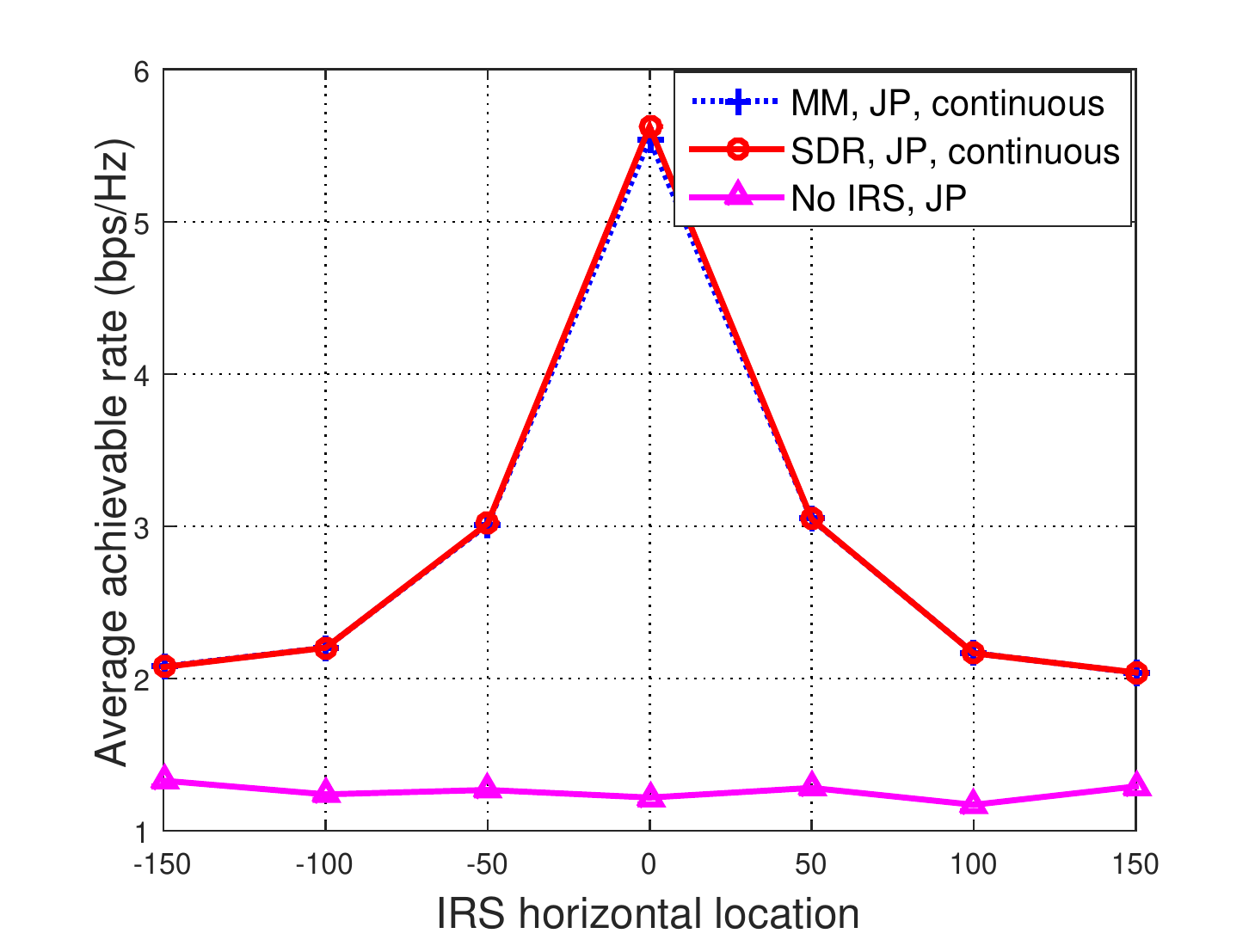}
	\caption{The impact of IRS location on the user's average achievable rate.}\label{fig6} 
\end{minipage}
\end{figure}

In order to show the performance gain brought by the IRS in the JP-CoMP transmission system, we compare the following schemes: 1) MM, JP, continuous:  our proposed Algorithm~\ref{alg3}; 2) SDR, JP: this  is realized by our proposed scheme, while  the difference from scheme 1) lies in  the phase shift matrix which is solved  based on the  SDR technique as discussed in Section IV for the case of multiuser; 3) Random  phase, JP: each phase shift  at the IRS is random and follows uniform distribution over $\left[ {0{\kern 1pt} {\kern 1pt} {\kern 1pt} {\kern 1pt} {\kern 1pt} 2\pi } \right)$ for each channel generation;  4) No IRS, JP: without using the IRS.  For scheme ``SDR, JP'', we further consider two resolutions of  phase shifts, namely  $b=1$ and $b=2$. In Fig.~\ref{fig3},  we compare the average achievable rate obtained by the above   schemes versus the  BS transmit power budget under $M=100$ and $N_t=2$.  It is observed that the average achievable rate obtained by all the schemes increases with the BS transmit power budget. Besides, both the ``MM, JP, continuous" and ``SDR, JP'' schemes outperform the ``No IRS, JP" scheme significantly, which demonstrates that the system performance can  indeed be improved significantly with the deployment of an IRS. It  is also  observed  that ``SDR, JP'' with discrete phase shifts suffers from some  performance losses compared to the ``SDR, JP, continuous'' scheme. However, the  performance loss can be compensated  by adopting  a high-resolution phase shifts. Furthermore, the IRS adopting  random phase shifts still outperforms the scheme without IRS, as the IRS is able to reflect some of the dissipated signals back to the desired users.  Finally, the ``MM, JP, continuous'' scheme achieves  nearly the same performance as  the ``SDR, JP, continuous'' scheme,  but  with  much lower computational complexity as discussed in Section III-C and Section IV-C.

In Fig.~\ref{fig5},  the average achievable rate obtained by all the schemes  versus the  number of IRS reflecting elements is studied.  It is observed that the proposed  schemes including ``SDR, JP'' and ``MM, JP, continuous'' outperform that of both the ``Random phase, JP'' scheme and  the  ``No IRS, JP'' scheme. Especially, for a larger  $M$, the system performance gain is more pronounced. For example, when $M=50$, the average rate achieved by the ``No IRS'' scheme is about 1.29$~\rm {bps/Hz}$ and that by  ``SDR, JP, continuous'' scheme is about 4.62$~\rm {bps/Hz}$, while when $M=300$, the latter increases up to 7.76$~\rm {bps/Hz}$.
This is because   installing more passive reflecting elements provides more degrees of freedom for resource allocation, which is beneficial for achieving higher beamforming gain, thereby improve the system throughput. More importantly,   since the IRS is passive with  low power consumption and low cost, it is promising for applying an IRS with  hundreds even thousands of  reflecting elements.

In Fig.~\ref{fig6}, we study the impact of  IRS's location  on the user's average achievable rate under $M=100$ and $N_t=2$. We assume that the IRS locates at right above the line  connecting  two BSs, and the horizontal axis  ranges from $\left[ {{\rm{ - }}150~{\rm{m}},150~{\rm{m}}} \right]$. It is observed  that as the IRS is deployed closer to the user, a higher rate can be achieved  due to the smaller reflection path loss. This indicates  the potential performance gain brought by the deployment of IRS, especially when IRS is close to the user.

\subsection{Multiuser System}
\begin{figure}[!t]
\centering
\begin{minipage}[t]{0.45\textwidth}
\centering
\includegraphics[width=2.5in]{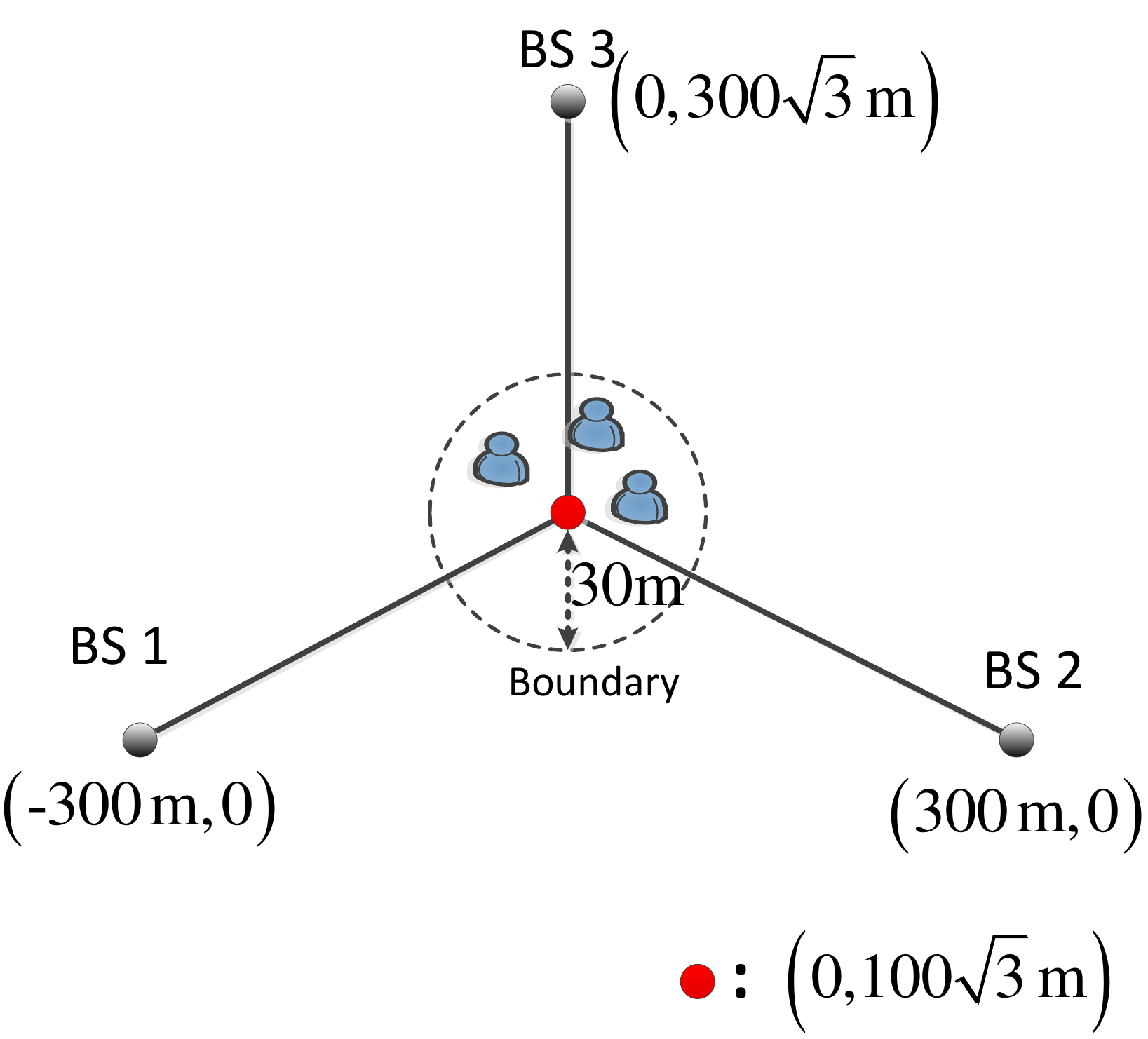}
\caption{Simulation setup for the multiuser system.}\label{fig6-1}
\end{minipage}
\begin{minipage}[t]{0.45\textwidth}
\centering
\includegraphics[width=2.8in]{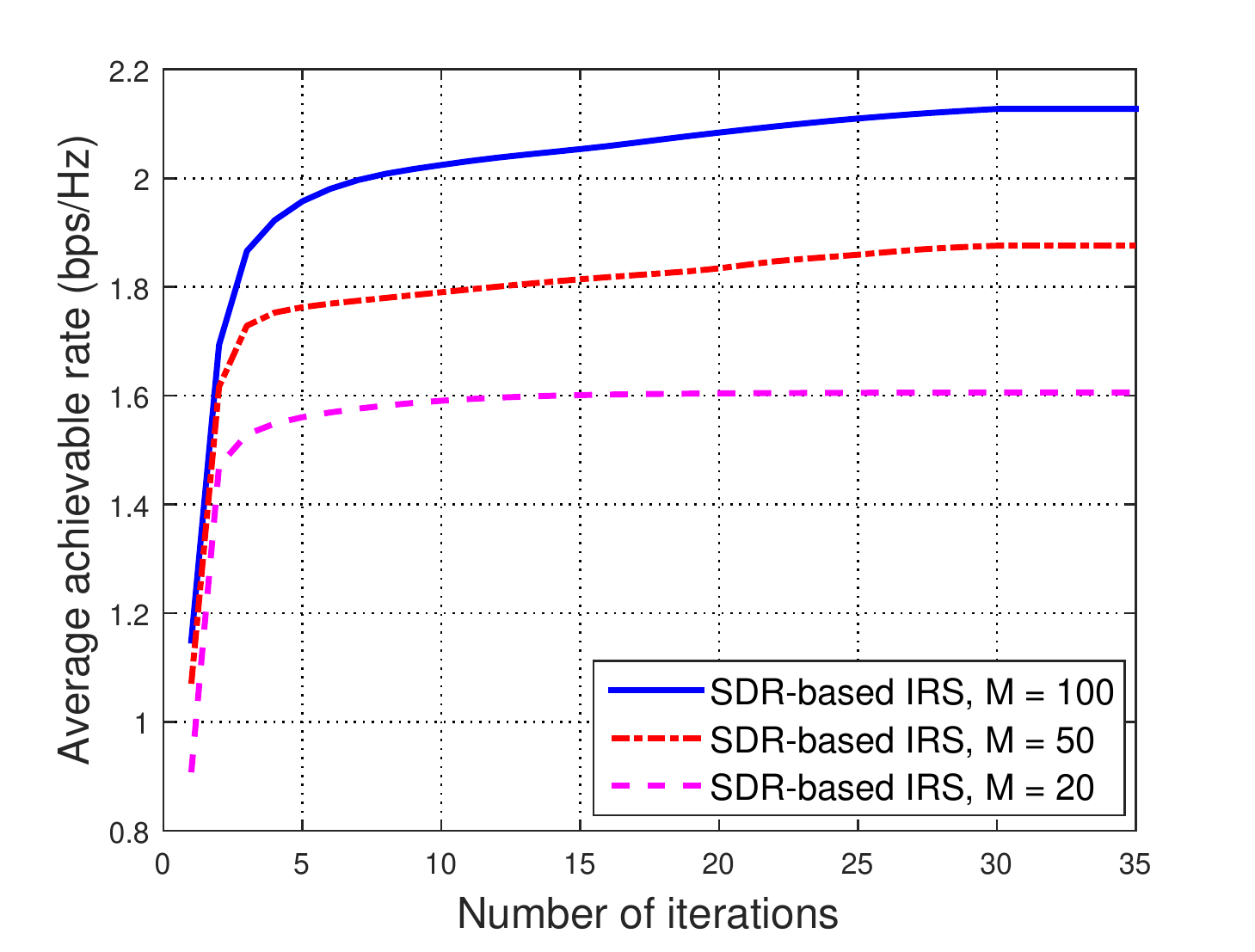}
\caption{Iteration behaviour of  Algorithm~\ref{alg4}.}\label{fig7}
\end{minipage}
\end{figure}
Next, we consider the  multiuser system, where there are three  BSs and  three  cell-edge users
as shown in Fig.~\ref{fig6-1}. The three  BSs are respectively located at $\left( {{\rm{ - }}300~{\rm{m}},0} \right)$, $\left( {300~{\rm{m}},0} \right)$, and $\left( {0,300\sqrt 3 ~{\rm{m}}} \right)$. Moreover, all the users are uniformly and randomly distributed in a circle centered at $\left( {0,100\sqrt 3 ~{\rm{m}}} \right)$ with a radius $30~\rm m$. The IRS is right above the  central point $\left( {0,100\sqrt 3 ~{\rm{m}}} \right)$ with  altitude $10~\rm m$. Unless otherwise stated, we set $N_t=6$, $N_r=2$,  $M=100$ for this scenario.

To show the efficiency  of proposed Algorithm~\ref{alg4},  its iteration behaviour  for the different numbers of the IRS reflecting elements   is plotted in Fig.~\ref{fig7}. It is observed that the  average max-min rate increases quickly with the number of iterations. In particular, for  $M=20$, the proposed algorithm terminates in about $5$ iterations, while for   $M=100$, only about $25$ iterations are required for reaching the termination.

\begin{figure}[!t]
	\centering
	\begin{minipage}[t]{0.45\textwidth}
		\centering
		\includegraphics[width=2.8in]{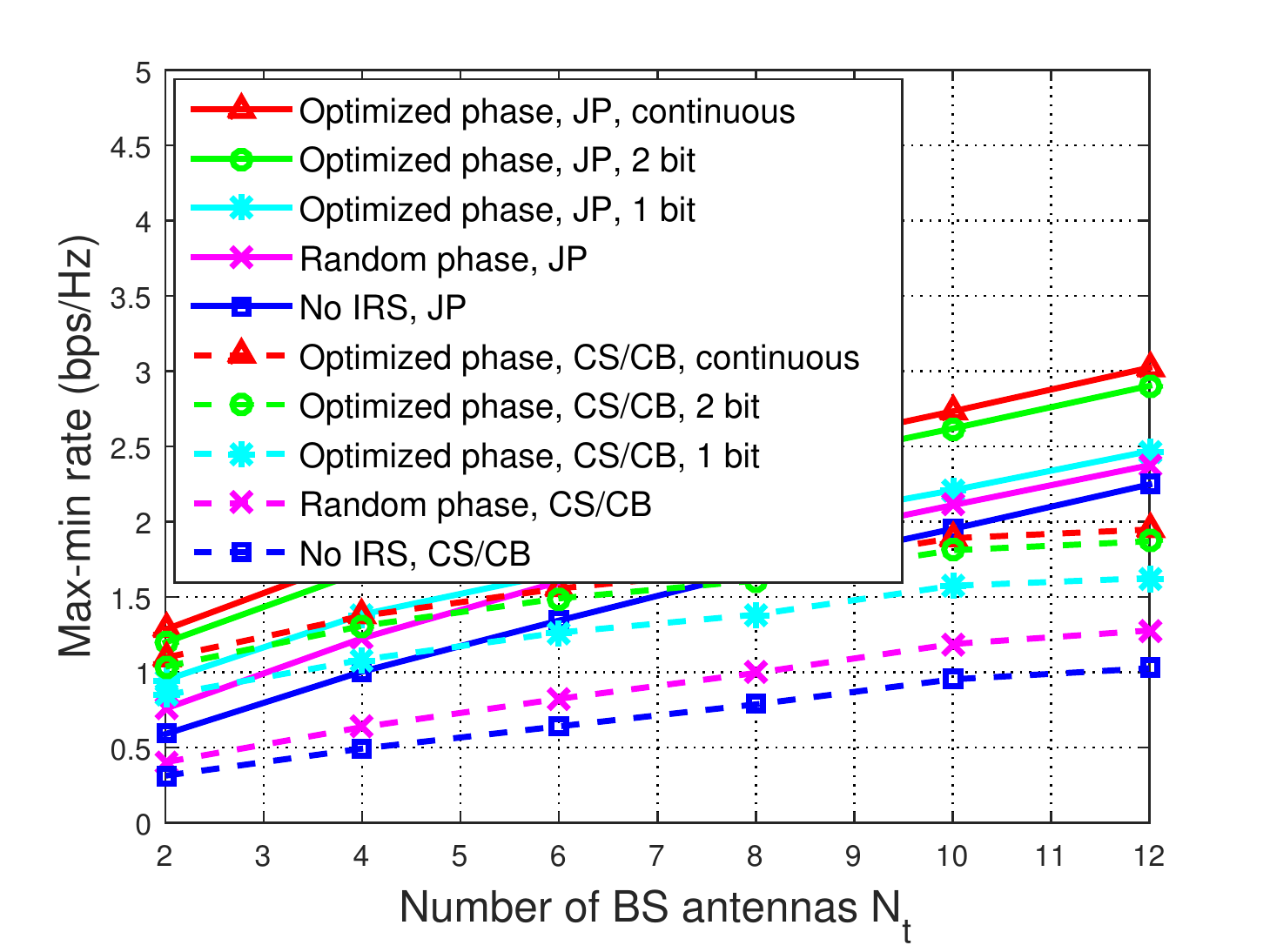}
		\caption{Average max-min rate versus  the number of BS antenna $N_t$.} \label{fig7-2}
	\end{minipage}
	\begin{minipage}[t]{0.45\textwidth}
	\centering
	\includegraphics[width=2.8in]{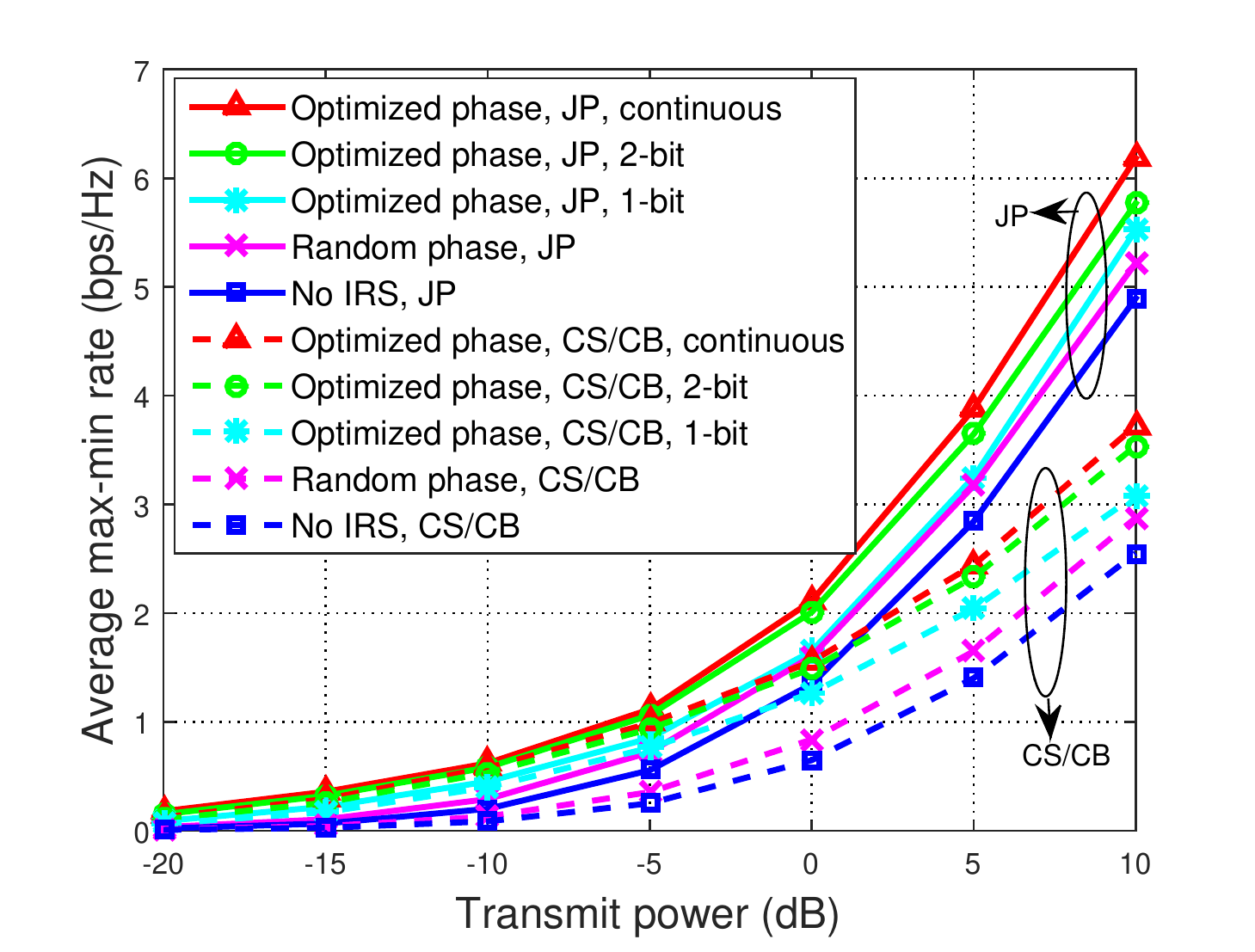}
	\caption{Average max-min rate versus  BS transmit power budget.}\label{fig8}
\end{minipage}
\end{figure}

In order to show the performance gain brought by the IRS-aided  JP-CoMP design, the IRS-aided CS/CB-CoMP design proposed in \cite{pan2019multicell} is  considered here for comparison. We compare the following schemes  for  the JP-CoMP  systems: 1) Optimized phase, JP: proposed Algorithm~\ref{alg4} in Section IV for the JP-CoMP  systems; 2) Random  phase, JP: each phase shift  at the IRS is randomly chosen from  the uniform distribution over $\left[ {0{\kern 1pt} {\kern 1pt} {\kern 1pt} {\kern 1pt} {\kern 1pt} 2\pi } \right)$ in each channel generation; 3) No  IRS, JP: without adopting the IRS in the  JP-CoMP systems.  Similarly, we have the  same  counterpart   schemes for the CS/CB-CoMP systems. Also as in Section V-A,  we consider both $b = 1$ and $b = 2$ for the discrete phase shifts at the IRS. In Fig.~\ref{fig7-2},  the average max-min  rate obtained by all the schemes versus number of BS antennas is plotted. It is observed that the average max-min  rate obtained by all the schemes increases with the number of BS antennas. One can observe that the proposed ``Optimized phase, JP, 2-bit'' scheme only suffers from a very small performance loss as  compared to that the scheme with continuous phase shifts. This shows the great benefits of array 
gain brought by the multiple number of antennas. In addition, we can also see that  our JP schemes outperform their counterpart CS/CB schemes in terms of average max-min rate as the proposed scheme can fully utilize the specific  degrees of freedom in the system for resource allocation.

In Fig.~\ref{fig8},  we compare the average max-min  rate versus the BS transmit power budget.   It is firstly observed from Fig.~\ref{fig8} that in terms of the average max-min data rate, the optimized phase schemes for the JP-CoMP design substantially outperform that for  the CS/CB-CoMP design, which demonstrates the superiority of the proposed IRS-aided JP-CoMP design. Particularly,  when the BS has a larger transmit power budget, the performance gap between two designs becomes more pronounced. This is because  inter-user interference can be  significantly suppressed by the JP-CoMP technique such that the resource allocation can fully exploit  a large transmit power budget at each BS to improve the average max-min rate. For example, when $P_{\rm max}=10~{\rm dB}$, the average max-min rate obtained by ``Optimized phase, CS/CB, continuous''  is   $3.7246~{\rm bps/Hz}$, and that obtained by  ``Optimized phase, JP, continuous''  is  $6.1837 ~{\rm bps/Hz}$, which shows a nearly $40\%$ increase.   Besides, one can observe that the average max-min rate of using discrete phase shifts is  significantly higher than that  without IRS for large transmit power at the BS, which demonstrates the advantage of optimizing phase shifts. We can also see that adopting the IRS with discrete phase shifts suffers some small performance losses compared to the IRS with continuous phase shifts for both designs.  This is expected since the  multi-path signals  cannot be perfectly aligned in phase at receivers in the case with  discrete phase shifters, thus resulting in some performance losses. However, with a higher resolution  IRS, i.e., $b=2$, the performance loss has been significantly reduced compared to the IRS with continuous phase shifts.

\begin{figure}[!t]
	\centering
	\begin{minipage}[t]{0.45\textwidth}
		\centering
		\includegraphics[width=2.8in]{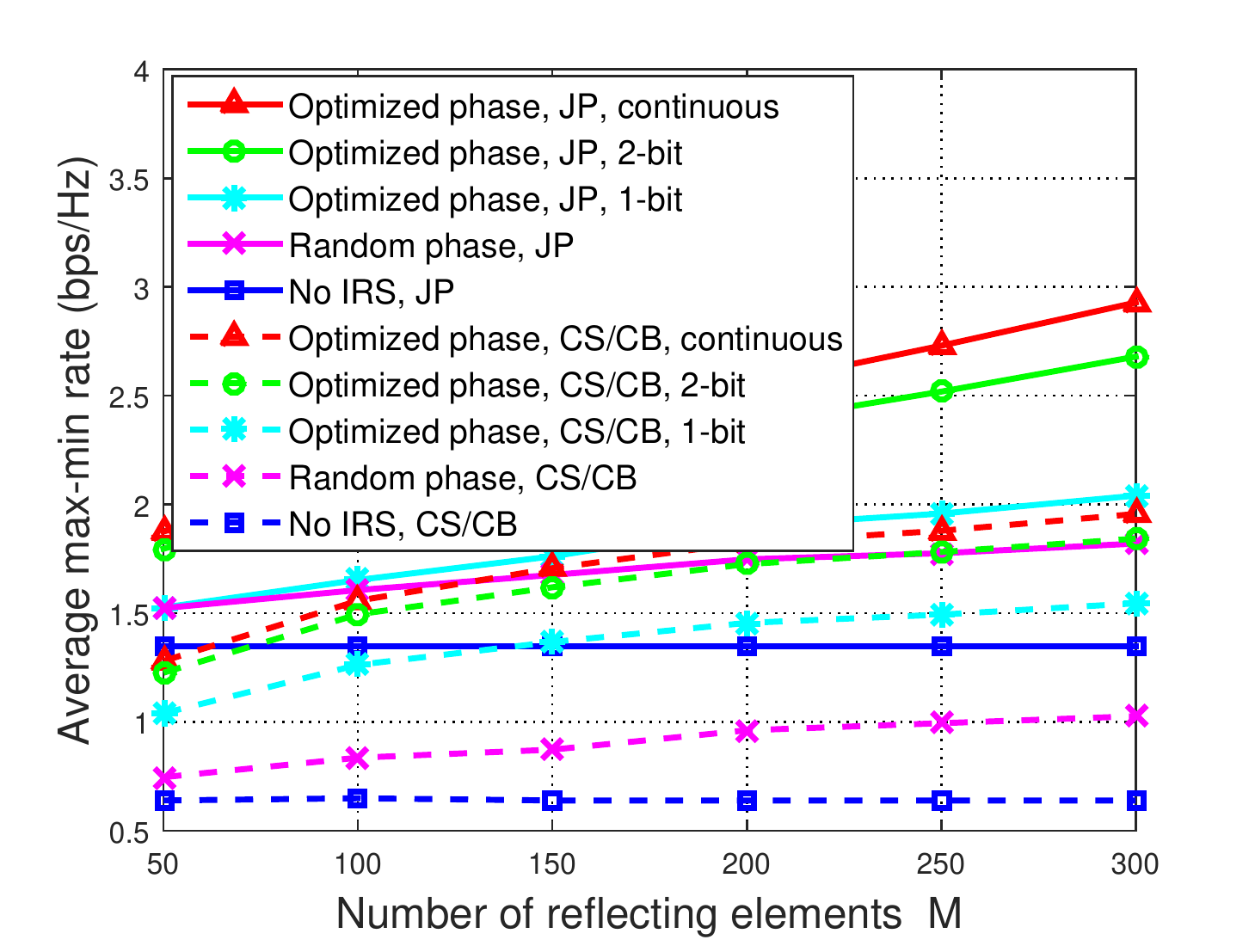}
		\caption{Average max-min rate versus  the number of IRS reflecting elements.} \label{fig10}
	\end{minipage}
	\begin{minipage}[t]{0.45\textwidth}
		\centering
		\includegraphics[width=2.8in]{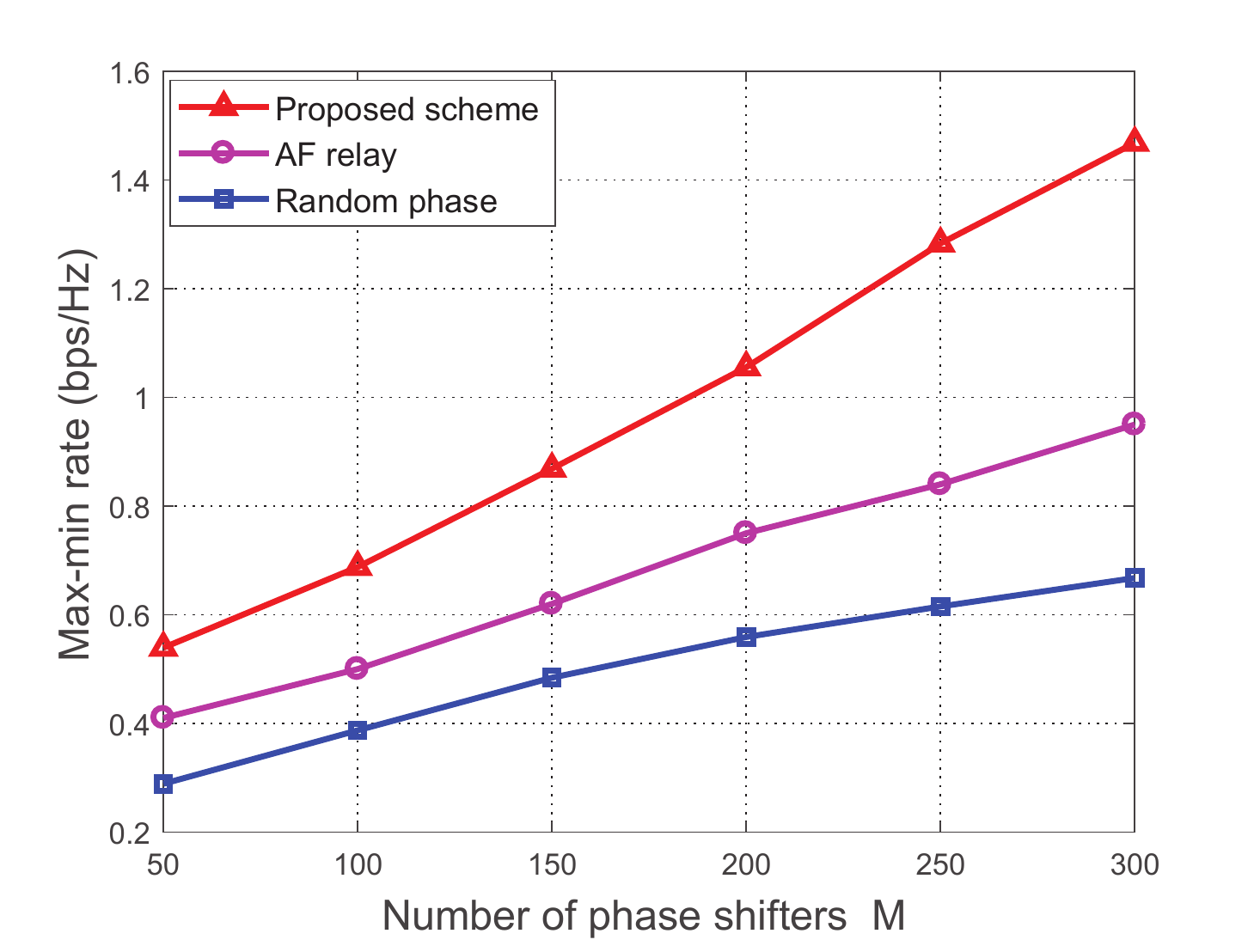}
		\caption{Comparison between the IRS and the  AF relay.}\label{fig11}
	\end{minipage}
\end{figure}

In Fig.~\ref{fig10}, the average max-min  rate versus the number of IRS reflecting elements is studied. It is observed that  the performance gain of the IRS-aided JP-CoMP  scheme  increases as the number of IRS reflecting elements increases, since  more  reflecting elements help achieve  higher passive beamforming gain. In addition, we can observe that the performance gap between ``Optimized phase, JP, continuous'' and ``No IRS, JP'' is magnified as $M$  increases. This is because  the   beam reflected by the IRS towards the desired users becomes more focused and powerful with increasing $M$.
This again shows that deploying IRS is a promising solution to address the network capacity bottleneck issue.
In addition, we can still observe  that the JP-CoMP design outperforms  the CS/CB-CoMP design in terms of average max-min rate, especially with large $M$, which further  demonstrates the superiority of our proposed JP-CoMP  design over the CS/CB-CoMP design. These findings also reinforce also  the motivation of our paper, that the combination of JP-CoMP and IRS provides symbiotic benefits to improve the system performance.

Finally, we compare the average max-min rate of the  IRS versus the half-duplex amplify-and-forward (AF) relay. 
	To focus on the comparison with large reflecting elements $M$, the direct link from
	the BS to the user is ignored. Note that  \cite{huang2019Reconfigurable}, \cite{wu2019intelligent}, \cite{ebj2020Intelligent}, and \cite{direnzo2020Reconfigurable} only focus a single-cell system with  single-antenna at the receiver, while our work focuses on a more complex multicell multi-antenna multiple user system. For the  half duplex AF  relaying protocol,  the transmission is divided into two equal-sized phases. In the first phase, the data  received by the AF relay is 
	\begin{align}
	{{\bf{y}}_r} = \sum\limits_{n = 1}^N {{{\bf{G}}_{n,r}}} {{\bf{x}}_n} + {{\bf{n}}_r}= \sum\limits_{j = 1}^K {{{\bf{G}}_r}{{\bf{W}}_j}} {{\bf{s}}_j} + {{\bf{n}}_r},
	\end{align}
	where ${{\bf{G}}_r} = \left[ {{{\bf{G}}_{1,r}}, \ldots ,{{\bf{G}}_{N,r}}} \right]$,  ${{{\bf{W}}_j}}$ is defined in Section II, and ${{\bf{n}}_r} \sim {\cal CN}\left( {{\bf{0}},{\sigma ^2}{{\bf{I}}_{{M}}}} \right)$ is the received noise.
	In the second phase, the relay transmits  data to the users, the received signal at user $k$ is given by 
	\begin{align}
	{\bf{y}}_r^{{\rm{AF}}} &= {{\bf{H}}_{r,k}}{\bf{V}}{{\bf{y}}_r} + {{\bf{n}}_k} \notag\\
	&= {{\bf{H}}_{r,k}}{\bf{V}}\sum\limits_{j = 1}^K {{{\bf{G}}_r}{{\bf{W}}_j}} {{\bf{s}}_j} + {{\bf{H}}_{r,k}}{\bf{V}}{{\bf{n}}_r} + {{\bf{n}}_k},
	\end{align}
	where ${\bf{V}} \in {{\mathbb C}^{M \times M}}$ is the relay processing matrix.  To compare with the IRS fairly, the analog beamforming matrix is considered, i.e., $\left( {i,j} \right)$th element of ${\bf{V}}$ satisfies $\left| {{{\bf{V}}_{i,j}}} \right| = 1$.
	
	As a result, the  achievable data rate (nat/s/Hz) of  user $k$ is given by
	\begin{align}
	R_k^{{\rm{AF}}} = \frac{1}{2}\ln \left| {{{\bf{I}}_{{N_r}}} + {{\bf{H}}_{r,k}}{\bf{V}}{{\bf{G}}_r}{{\bf{W}}_k}{\bf{W}}_k^H{\bf{G}}_r^H{{\bf{V}}^H}{\bf{H}}_{r,k}^H{\bf{F}}_{2,k}^{ - 1}} \right|,
	\end{align}
	where ${{\bf{F}}_{2,k}} = {{\bf{H}}_{r,k}}{\bf{V}}\left( {{{\bf{G}}_r}\sum\limits_{j \ne k}^K {{{\bf{W}}_j}{\bf{W}}_j^H} {\bf{G}}_r^H + {\sigma ^2}{{\bf{I}}_M}} \right){{\bf{V}}^H}{\bf{H}}_{r,k}^H + {\sigma ^2}{{\bf{I}}_M}$, the factor  $\frac{1}{2}$ denotes the half-duplex protocol  adopted  at the relay.
	As such, we have the following problem for the half-duplex AF relaying systems:
	\begin{align}
	\left( {\rm{P}3} \right):&\mathop {\rm maximize }\limits_{{{\bf{W}}_{n,k}},{\bf V },R} {\kern 1pt} {\kern 1pt} {\kern 1pt} {\kern 1pt} R\notag\\
	&{\rm s.t.}~\frac{1}{2}\ln \left| {{{\bf{I}}_{{N_r}}} + {{\bf{H}}_{r,k}}{\bf{V}}{{\bf{G}}_r}{{\bf{W}}_k}{\bf{W}}_k^H{\bf{G}}_r^H{{\bf{V}}^H}{\bf{H}}_{r,k}^H{\bf{F}}_{2,k}^{ - 1}} \right| \notag\\
	&\qquad\qquad\qquad\qquad\qquad\qquad\qquad\ge R,k \in \cal {\cal K},\\
	&\qquad\sum\limits_{k = 1}^K {\left\| {{{\bf{W}}_{n,k}}} \right\|_F^2}  \le {P_{\max }}, {\kern 1pt} {\kern 1pt}n \in \cal N,\\
	&\qquad \left| {{{\bf{V}}_{i,j}}} \right| = 1,i \in {\cal M},j \in {\cal M}.
	\end{align}
Problem $({\rm P}3)$ is different from $({\rm P}3)$, where the non-diagonal element of ${\bf{V}}$ is constrained to  modulus $1$. To solve Problem $({\rm P}3)$, MSE and  alternating optimization method are  still applied. Specifically, for the given analog beamforming ${\bf{V}}$, the BS beamforming ${{{\bf{W}}_{n,k}}}$ can be solved similar to Section IV-A. As for the given BS beamforming ${{{\bf{W}}_{n,k}}}$, analog beamforming ${\bf{V}}$ can be solved by  updating  each phase shift sequentially according to \cite{wu2019beamforming}.  In Fig.~\ref{fig11}, we plot the 
average max-min  rate versus the number of IRS reflecting elements for the IRS and the AF relay. It is observed that proposed scheme (i.e., optimize IRS phase shift) outperforms  that of AF relaying. Specifically, 
when the number of reflecting elements $M$ is small, the performance gap between proposed scheme and AF relaying is small. However,  when $M$ is large, the  performance gap becomes significant. Since the IRS is passive, the signal impinging on IRS is directly reflected with no time delay due to its passive property, while the signal impinges on  relay, a time slot delay is induced due to the  active components applied at relay.

\section{Conclusion}
In this paper, we  studied the IRS-aided  JP-CoMP downlink transmission in  multi-cell  systems. To guarantee the user fairness, a max-min rate  problem was  formulated  by jointly optimizing the IRS phase shift matrix and the BS transmit beamforming matrix. We  considered both the  single cell-edge user system and multi-user system. For the single-user system, the transmit  beamforming matrix was optimally obtained based on the  subgradient method for the fixed IRS phase shift matrix, and the IRS phase shift matrix was  obtained  based on  the MM method for  the fixed transmit  beamforming matrix. Exploiting these two solutions, an efficient iterative resource allocation algorithm  was proposed. For the multi-user system, with the given phase shift matrix, the transmit beamforming  optimization problem was transformed into an SOCP, which was efficiently solved by the interior point method. For the given  transmit beamforming matrix, the IRS phase shift matrix  was optimized  by leveraging the SDR technique. Then, an efficient iterative algorithm was  also proposed.  Simulation results demonstrated that with the deployment of IRS, significant throughput  can be achieved over the case   without  IRS. Furthermore,  the  proposed JP-CoMP design significantly outperforms the CS/CB-CoMP design in terms of max-min rate.
The results in this paper can be further extended by considering multiple IRSs, frequency-selective channel model, imperfect CSI, etc., which will be left as future work. 
%\appendices
%\newpage
\begin{center}
{{\normalsize A}{\small PPENDIX} A} \label{appendix1}
\end{center}
%\section{Proof of Theorem~1 } \label{appendix1}

We prove the theorem  by examing the Karush-Kuhn-Tucker (KKT) conditions of both $(\rm P)$ and $(\rm P1)$. We first show that the KKT conditions of problem $\left( {\rm{P}} \right)$ is the same as that of $\left( {\rm{P1}} \right)$. To start with,  the Lagrangian function associated with constraint \eqref{P1const1} of $\left( {\rm{P1}} \right)$ is given by
\begin{align}
&{\cal L}\left( {{{\bf{W}}_{n,k}},{\theta _m},R,{{\bf{U}}_k},{{\bf{Q}}_k},{\lambda _k}} \right) = R\notag\\
& + \sum\limits_{k = 1}^K {{\lambda _k}} \left( {\ln \left| {{{\bf{Q}}_k}} \right| - {\rm{Tr}}\left( {{{\bf{Q}}_k}{{\bf{E}}_k}} \right) + d-R} \right), \label{appendix1const1}
\end{align}
where $\lambda _k\ge0$, $\forall k$, is the dual variable corresponding to  constraint \eqref{P1const1}. According to \cite{boyd2004convex}, all the  locally optimal solutions (including the globally optimal solutions) must satisfy the KKT conditions.  Specifically, by  setting  the first-order derivative of $\cal L$ with respect to variables ${{{\bf{Q}}_k}}$ and ${{\bf U}_k}$ to zero, we  have
\begin{align}
{\nabla _{{{\bf{Q}}_k}}}{\cal L}\left( {{{\bf{W}}_{n,k}},{\theta _m},R,{{\bf{U}}_k},{{\bf{Q}}_k},{\lambda _k}} \right) = {\bf 0}, \label{appendix1const2}
\end{align}
\begin{align}
{\nabla _{{{\bf{U}}_k}}}{\cal L}\left( {{{\bf{W}}_{n,k}},{\theta _m},R,{{\bf{U}}_k},{{\bf{Q}}_k},{\lambda _k}} \right) ={ \bf 0}, \label{appendix1const3}
\end{align}
respectively. Based on \eqref{appendix1const2} and \eqref{appendix1const3}, the optimal solutions  ${{{\bf{Q}}_k^{\rm opt}}}$ and ${{{\bf{U}}_k^{\rm opt}}}$ can be  derived as
\begin{align}
{\bf{Q}}_k^{\rm opt} = {\bf{E}}_k^{ - 1}, \label{appendix1const4}
\end{align}
\begin{align}
{\bf{U}}_k^{\rm opt} = {\left( {{{{\bf{\bar H}}}_k}\left( {\sum\limits_{j = 1}^K {{{\bf{W}}_j}{\bf{W}}_j^H} } \right){\bf{\bar H}}_k^H + {\sigma ^2}{\bf{I}}_{N_r}} \right)^{ - 1}}{{{\bf{\bar H}}}_k}{{\bf{W}}_k}. \label{appendix1const5}
\end{align}
Substituting \eqref{appendix1const5} into  \eqref{Pconst5}, the minimum MSE (MMSE) matrix is given by
\begin{align}
{\bf{E}}_k^{\rm mmse} = {\bf{I}}_d - {\bf{W}}_k^H{\bf{\bar H}}_k^H{\bf{J}}_k^{ - 1}{{{\bf{\bar H}}}_k}{{\bf{W}}_k}, \label{appendix1const6}
\end{align}
where ${{\bf{J}}_k} = \left( {{{{\bf{\bar H}}}_k}\left( {\sum\limits_{j = 1}^K {{{\bf{W}}_j}{\bf{W}}_j^H} } \right){\bf{\bar H}}_k^H + {\sigma ^2}{\bf{I}}_{N_r}} \right)$.
Then, substituting \eqref{appendix1const4} and \eqref{appendix1const5} into \eqref{appendix1const1}, we arrive at
\begin{align}
{\cal L}\left( {{{\bf{W}}_{n,k}},{\theta _m},R,{\lambda _k}} \right) = R + \sum\limits_{k = 1}^K {{\lambda _k}} \left( {\ln \left| {{{\left( {{\bf{E}}_k^{\rm mmse}} \right)}^{ - 1}}} \right|}-R \right). \label{appendix1const7}
\end{align}
Using the Duncan-Guttman formula ${\left( {{\bf{A}} - {\bf{UD^{-1}V}}} \right)^{ - 1}} = {{\bf{A}}^{ - 1}} + {{\bf{A}}^{ - 1}}{\bf{U}}{\left( {{\bf{D}} - {\bf{V}}{{\bf{A}}^{ - 1}}{\bf{U}}} \right)^{ - 1}}{\bf{V}}{{\bf{A}}^{ - 1}}$ \cite{zhang2017matrix}, we can rewrite \eqref{appendix1const7} as
\begin{align}
&{\cal L}\left( {{{\bf{W}}_{n,k}},{\theta _m},R,{\lambda _k}} \right) = R\notag\\
&+\sum\limits_{k = 1}^K {{\lambda _k}} \left( {\ln \left| {{\bf{I}}_{N_r} + {{{\bf{\bar H}}}_k}{{\bf{W}}_k}{\bf{W}}_k^H{\bf{\bar H}}_k^H{\bf{F}}_k^{ - 1}} \right|}-R \right). \label{appendix1const8}
\end{align}
It can be seen  that \eqref{appendix1const8} is also  a Lagrangian function of $\left( {\rm{P}} \right)$ provided that $\lambda_k$ is  the Lagrangian multipliers for  constraint  \eqref{Pconst1}. This indicates that  problems $\left( {\rm{P}} \right)$ and $\left( {\rm{P1}} \right)$ have the  same optimal primal solution, which thus  completes the proof of Theorem 1.
\vspace{2em}
\begin{center}
{{\normalsize A}{\small PPENDIX} B} \label{appendix2}
\end{center}
%\section{Proof of Theorem~2 } \label{appendix2}

With the identity ${\rm{Tr}}\left( {{\bf{AB}}} \right) = {\left( {{\rm{vec}}\left( {\bf{A}} \right)} \right)^H}{\rm{vec}}\left( {\bf{B}} \right)$ \cite{zhang2017matrix}, we have   $\left\| {{{\bf{W}}_{n,k}}} \right\|_F^2{\rm{ = Tr}}\left( {{\bf{W}}_{n,k}^H{{\bf{W}}_{n,k}}} \right) = {\left( {{\rm{vec}}\left( {{{\bf{W}}_{n,k}}} \right)} \right)^H}{\rm{vec}}\left( {{{\bf{W}}_{n,k}}} \right)$. Therefore, we can rewrite \eqref{Pconst2} as
\begin{align}
{\left\| {{{\bm{\eta }}_n}} \right\|_2} \le \sqrt {{P_{\max }}} ,\forall n, \label{appendix2const1}
\end{align}
where ${{{\bm{\eta }}_n}}$ is given by
\begin{align}
{{\bm{\eta }}_n} = {\left[ {{{\left( {{\rm{vec}}\left( {{{\bf{W}}_{n,1}}} \right)} \right)}^H}, \cdots ,{{\left( {{\rm{vec}}\left( {{{\bf{W}}_{n,K}}} \right)} \right)}^H}} \right]^H}. \label{appendix2const2}
\end{align}
For constraint \eqref{P1const1}, we first rewrite ${{\bf{E}}_k}$ in \eqref{Pconst5} as
\begin{align}
&{{\bf{E}}_k} = \left( {{\bf{U}}_k^H{{{\bf{\bar H}}}_k}{{\bf{W}}_k} - {\bf{I}}} \right){\left( {{\bf{U}}_k^H{{{\bf{\bar H}}}_k}{{\bf{W}}_k} - {\bf{I}}} \right)^H} \notag\\
&+{\bf{U}}_k^H{{{\bf{\bar H}}}_k}\left( {\sum\limits_{j \ne k}^K {{{\bf{W}}_j}{\bf{W}}_j^H} } \right){\bf{\bar H}}_k^H{{\bf{U}}_k} + {\sigma ^2}{\bf{U}}_k^H{{\bf{U}}_k}.\label{appendix2const3}
\end{align}
Substituting \eqref{appendix2const3} into \eqref{P1const1}, we have
\begin{align}
{\left\| {{{\bm{\omega }}_k}} \right\|_2} \le \sqrt {\ln \left| {{{\bf{Q}}_k}} \right| + d - R - {\sigma ^2}{\rm{Tr}}\left( {{{\bf{Q}}_k}{\bf{U}}_k^H{{\bf{U}}_k}} \right)} ,\forall k, \label{appendix2const4}
\end{align}
where ${{\bm{\omega }}_k}$ is given by
\begin{align}
%{{\bm{\omega }}_k} = {\left[ {{{\left( {{\rm{vec}}\left( {{\bf{W}}_1^H{\bf{\bar H}}_k^H{{\bf{U}}_k}{\bf{Q}}_k^{1/2}} \right)} \right)}^H}, \cdots ,{{\left( {{\rm{vec}}\left( {{\bf{W}}_k^H{\bf{\bar H}}_k^H{{\bf{U}}_k}{\bf{Q}}_k^{1/2} - {\bf{Q}}_k^{1/2}} \right)} \right)}^H}, \cdots ,{{\left( {{\rm{vec}}\left( {{\bf{W}}_K^H{\bf{\bar H}}_k^H{{\bf{U}}_k}{\bf{Q}}_k^{1/2}} \right)} \right)}^H}} \right]^H}. \label{appendix2const5}
{{\bm{\omega }}_k}& = \left[ {{{\left( {{\rm{vec}}\left( {{\bf{W}}_1^H{\bf{\bar H}}_k^H{{\bf{U}}_k}{\bf{Q}}_k^{1/2}} \right)} \right)}^H}} \right.\notag\\
&, \cdots ,{\left( {{\rm{vec}}\left( {{\bf{W}}_k^H{\bf{\bar H}}_k^H{{\bf{U}}_k}{\bf{Q}}_k^{1/2} - {\bf{Q}}_k^{1/2}} \right)} \right)^H}\notag\\
&, \cdots ,{\left. {{{\left( {{\rm{vec}}\left( {{\bf{W}}_K^H{\bf{\bar H}}_k^H{{\bf{U}}_k}{\bf{Q}}_k^{1/2}} \right)} \right)}^H}} \right]^H}.\label{appendix2const5}
% {\bm {\omega _k}} = &\left[ {{{\left( {{\rm{vec}}\left( {{\bf{W}}_1^H{\bf{\bar H}}_k^H{{\bf{U}}_k}{\bf{Q}}_k^{1/2}} \right)} \right)}^H}, \cdots ,{{\left( {{\rm{vec}}\left( {{\bf{W}}_k^H{\bf{\bar H}}_k^H{{\bf{U}}_k}{\bf{Q}}_k^{1/2} - {\bf{Q}}_k^{1/2}} \right)} \right)}^H},} \right.\notag\\
%& \cdots ,{\left. {{{\left( {{\rm{vec}}\left( {{\bf{W}}_K^H{\bf{\bar H}}_k^H{{\bf{U}}_k}{\bf{Q}}_k^{1/2}} \right)} \right)}^H}} \right]^H}. \label{appendix2const5}
\end{align}

This  completes the proof of Theorem 2.
\bibliographystyle{IEEEtran}
\bibliography{COMPIRS}

\begin{thebibliography}{10}
\providecommand{\url}[1]{#1}
\csname url@samestyle\endcsname
\providecommand{\newblock}{\relax}
\providecommand{\bibinfo}[2]{#2}
\providecommand{\BIBentrySTDinterwordspacing}{\spaceskip=0pt\relax}
\providecommand{\BIBentryALTinterwordstretchfactor}{4}
\providecommand{\BIBentryALTinterwordspacing}{\spaceskip=\fontdimen2\font plus
\BIBentryALTinterwordstretchfactor\fontdimen3\font minus
  \fontdimen4\font\relax}
\providecommand{\BIBforeignlanguage}[2]{{%
\expandafter\ifx\csname l@#1\endcsname\relax
\typeout{** WARNING: IEEEtran.bst: No hyphenation pattern has been}%
\typeout{** loaded for the language `#1'. Using the pattern for}%
\typeout{** the default language instead.}%
\else
\language=\csname l@#1\endcsname
\fi
#2}}
\providecommand{\BIBdecl}{\relax}
\BIBdecl

\bibitem{lu2014overview}
L.~Lu, G.~Y. Li, A.~L. Swindlehurst, A.~Ashikhmin, and R.~Zhang, ``An overview
  of massive {MIMO}: Benefits and challenges,'' \emph{{IEEE} J. Sel. Top. Sign.
  Proces.}, vol.~8, no.~5, pp. 742--758, Oct. 2014.

\bibitem{larsson2014massive}
E.~G. Larsson, O.~Edfors, F.~Tufvesson, and T.~L. Marzetta, ``Massive {MIMO}
  for next generation wireless systems,'' \emph{IEEE Commun. Mag.}, vol.~52,
  no.~2, pp. 186--195, Feb. 2014.

\bibitem{swindlehurst2014millimeter}
A.~L. Swindlehurst, E.~Ayanoglu, P.~Heydari, and F.~Capolino, ``Millimeter-wave
  massive {MIMO}: The next wireless revolution?'' \emph{{IEEE} Commun. Mag.},
  vol.~52, no.~9, pp. 56--62, Sep. 2014.

\bibitem{kamel2016ultra}
M.~Kamel, W.~Hamouda, and A.~Youssef, ``Ultra-dense networks: {A} survey,''
  \emph{IEEE Commun. Surveys Tuts.}, vol.~18, no.~4, pp. 2522--2545, 4th Quat.
  2016.

\bibitem{wong2017key}
V.~W. Wong, D.~W.~K. Ng, R.~Schober, and {L.-C. Wang}, \emph{Key technologies
  for {5G} wireless systems}.\hskip 1em plus 0.5em minus 0.4em\relax Cambridge
  university press, 2017.

\bibitem{2020Massivechen}
\BIBentryALTinterwordspacing
X.~Chen, D.~W.~K. Ng, W.~Yu, E.~G. Larsson, N.~Al-Dhahir, and R.~Schober,
  ``Massive access for {5G} and beyond,'' 2020. [Online]. Available:
  \url{https://arxiv.org/abs/2002.03491.}
\BIBentrySTDinterwordspacing

\bibitem{zhang2017fundamental}
S.~{Zhang}, Q.~{Wu}, S.~{Xu}, and G.~Y. {Li}, ``Fundamental green tradeoffs:
  Progresses, challenges, and impacts on {5G} networks,'' \emph{IEEE Commun.
  Surveys Tuts.}, vol.~19, no.~1, pp. 33--56, 1st Quat. 2017.

\bibitem{wu2017an}
Q.~{Wu}, G.~Y. {Li}, W.~{Chen}, D.~W.~K. {Ng}, and R.~{Schober}, ``An overview
  of sustainable green {5G} networks,'' \emph{IEEE Wireless Commun.}, vol.~24,
  no.~4, pp. 72--80, Aug. 2017.

\bibitem{wu2020towards}
Q.~{Wu} and R.~{Zhang}, ``Towards smart and reconfigurable environment:
  Intelligent reflecting surface aided wireless network,'' \emph{{IEEE} Commun.
  Mag.}, vol.~58, no.~1, pp. 106--112, Jan. 2020.

\bibitem{basar2019wireless}
E.~Basar, M.~Di~Renzo, J.~De~Rosny, M.~Debbah, M.-S. Alouini, and R.~Zhang,
  ``Wireless communications through reconfigurable intelligent surfaces,''
  \emph{IEEE Access}, vol.~7, pp. 116\,753--116\,773, 2019.

\bibitem{zhao2019survey}
\BIBentryALTinterwordspacing
J.~Zhao, ``A survey of intelligent reflecting surfaces ({IRSs}): Towards {6G}
  wireless communication networks with massive {MIMO} 2.0.'' [Online].
  Available: \url{https://arxiv.org/abs/1907.04789.}
\BIBentrySTDinterwordspacing

\bibitem{zhang2019multiple}
J.~Zhang, E.~Bj{\"o}rnson, M.~Matthaiou, D.~W.~K. Ng, H.~Yang, and D.~J. Love,
  ``Prospective multiple antenna technologies for beyond {5G},'' \emph{IEEE J.
  Sel. Areas Commun.}, vol.~38, no.~8, pp. 1637--1660, Aug. 2020.

\bibitem{cui2017information}
T.~J. Cui, S.~Liu, and L.~Zhang, ``Information metamaterials and
  metasurfaces,'' \emph{J. Phys. Chem. C}, vol.~5, no.~15, pp. 3644--3668,
  2017.

\bibitem{cui2014coding}
T.~J. Cui, M.~Q. Qi, X.~Wan, J.~Zhao, and Q.~Cheng, ``Coding metamaterials,
  digital metamaterials and programmable metamaterials,'' \emph{Light Sci.
  Appl.}, vol.~3, no.~10, p. e218, Oct. 2014.

\bibitem{Nayeri2018Reflectarray}
P.~Nayeri, F.~Yang, and A.~Z. Elsherbeni, \emph{Reflectarray Antennas: Theory,
  Designs, and Applications}.\hskip 1em plus 0.5em minus 0.4em\relax John Wiley
  \& Sons, 2018.

\bibitem{Liaskos2018Realizing}
C.~{Liaskos}, S.~{Nie}, A.~{Tsioliaridou}, A.~{Pitsillides}, S.~{Ioannidis},
  and I.~{Akyildiz}, ``Realizing wireless communication through
  software-defined hypersurface environments,'' in \emph{2018 IEEE 19th
  International Symposium on "A World of Wireless, Mobile and Multimedia
  Networks" (WoWMoM)}, 2018, pp. 14--15.

\bibitem{ma2019smart}
Q.~Ma, G.~D. Bai, H.~B. Jing, C.~Yang, L.~Li, and T.~J. Cui, ``Smart
  metasurface with self-adaptively reprogrammable functions,''
  \emph{Light-Science \& Applications}, vol.~8, no.~1, pp. 1--12, 2019.

\bibitem{Liaskos2018anew}
C.~{Liaskos}, S.~{Nie}, A.~{Tsioliaridou}, A.~{Pitsillides}, S.~{Ioannidis},
  and I.~{Akyildiz}, ``A new wireless communication paradigm through
  software-controlled metasurfaces,'' \emph{IEEE Commun. Mag.}, vol.~56, no.~9,
  pp. 162--169, 2018.

\bibitem{huang2020Holographic}
C.~{Huang}, S.~{Hu}, G.~C. {Alexandropoulos}, A.~{Zappone}, C.~{Yuen},
  R.~{Zhang}, M.~{Di Renzo}, and M.~{Debbah}, ``Holographic mimo surfaces for
  {6G} wireless networks: Opportunities, challenges, and trends,'' \emph{IEEE
  Wireless Commun.}, 2020, early access, doi: 10.1109/MWC.001.1900534.

\bibitem{huang2019Reconfigurable}
C.~{Huang}, A.~{Zappone}, G.~C. {Alexandropoulos}, M.~{Debbah}, and C.~{Yuen},
  ``Reconfigurable intelligent surfaces for energy efficiency in wireless
  communication,'' \emph{IEEE Trans. Wireless Commun.}, vol.~18, no.~8, pp.
  4157--4170, 2019.

\bibitem{wu2019intelligent}
Q.~Wu and R.~Zhang, ``Intelligent reflecting surface enhanced wireless network
  via joint active and passive beamforming,'' \emph{IEEE Trans. Wireless
  Commun.}, vol.~18, no.~11, pp. 5394--5409, Nov. 2019.

\bibitem{Zhu2013Active}
B.~O. Zhu, J.~Zhao, and Y.~Feng, ``Active impedance metasurface with full 360
  reflection phase tuning,'' \emph{Scientific Reports}, vol.~3, 2013.

\bibitem{Cui2016Information}
T.-J. Cui, S.~Liu, and L.-L. Li, ``Information entropy of coding metasurface,''
  \emph{Light Science \& Applications}, vol.~5, no.~11, p. e16172, 2016.

\bibitem{Abeywickrama2020Intelligent}
S.~{Abeywickrama}, R.~{Zhang}, Q.~{Wu}, and C.~{Yuen}, ``Intelligent reflecting
  surface: Practical phase shift model and beamforming optimization,''
  \emph{IEEE Trans. Commun.}, vol.~68, no.~9, pp. 5849--5863, Sept. 2020.

\bibitem{tang2019wireless}
W.~{Tang}, M.~Z. {Chen}, X.~{Chen}, J.~Y. {Dai}, Y.~{Han}, M.~{Di Renzo},
  Y.~{Zeng}, S.~{Jin}, Q.~{Cheng}, and T.~J. {Cui}, ``Wireless communications
  with reconfigurable intelligent surface: Path loss modeling and experimental
  measurement,'' \emph{IEEE Trans. Wireless Commun.}, 2020, doi:
  10.1109/TWC.2020.3024887.

\bibitem{you2019intelligent}
C.~{You}, B.~{Zheng}, and R.~{Zhang}, ``Intelligent reflecting surface with
  discrete phase shifts: Channel estimation and passive beamforming,'' in
  \emph{IEEE International Conference on Communications (ICC)}, 2020, pp. 1--6.

\bibitem{mirza2019channel}
\BIBentryALTinterwordspacing
J.~Mirza and B.~Ali, ``Channel estimation method and phase shift design for
  reconfigurable intelligent surface assisted {MIMO} networks.'' [Online].
  Available: \url{https://arxiv.org/abs/1912.10671.}
\BIBentrySTDinterwordspacing

\bibitem{chen2019channel}
\BIBentryALTinterwordspacing
J.~Chen, Y.-C. Liang, H.~V. Cheng, and W.~Yu, ``Channel estimation for
  reconfigurable intelligent surface aided multi-user {MIMO} systems.''
  [Online]. Available: \url{https://arxiv.org/abs/1912.03619.}
\BIBentrySTDinterwordspacing

\bibitem{wu2019beamforming}
Q.~{Wu} and R.~{Zhang}, ``Beamforming optimization for wireless network aided
  by intelligent reflecting surface with discrete phase shifts,'' \emph{IEEE
  Trans. Commun.}, vol.~68, no.~3, pp. 1838--1851, Mar. 2020.

\bibitem{pan2019multicell}
C.~Pan, H.~Ren, K.~Wang, W.~Xu, M.~Elkashlan, A.~Nallanathan, and L.~Hanzo,
  ``Multicell {MIMO} communications relying on intelligent reflecting
  surface,'' \emph{IEEE Trans. Wireless Commun.}, vol.~19, no.~8, pp.
  5218--5233, 2020.

\bibitem{pan2019intelligent}
C.~Pan, H.~Ren, K.~Wang, M.~Elkashlan, A.~Nallanathan, J.~Wang, and L.~Hanzo,
  ``Intelligent reflecting surface enhanced {MIMO} broadcasting for
  simultaneous wireless information and power transfer,'' \emph{IEEE J. Sel.
  Areas Commun.}, vol.~38, no.~8, pp. 1719--1734, Aug. 2020.

\bibitem{liu2019joint}
\BIBentryALTinterwordspacing
R.~Liu, M.~Li, Q.~Liu, and A.~L. Swindlehurst, ``Joint symbol-level precoding
  and reflecting designs for {RIS}-enhanced {MU-MISO} systems.'' [Online].
  Available: \url{https://arxiv.org/abs/1912.11767.}
\BIBentrySTDinterwordspacing

\bibitem{huang2018energy}
C.~{Huang}, G.~C. {Alexandropoulos}, A.~{Zappone}, M.~{Debbah}, and C.~{Yuen},
  ``Energy efficient multi-user {MISO} communication using low resolution large
  intelligent surfaces,'' in \emph{2018 IEEE Globecom Workshops (GC Wkshps)},
  2018, pp. 1--6.

\bibitem{wu2019weighted}
Q.~{Wu} and R.~{Zhang}, ``Weighted sum power maximization for intelligent
  reflecting surface aided {SWIPT},'' \emph{IEEE Wireless Commun. Lett.},
  vol.~9, no.~5, pp. 586--590, May 2020.

\bibitem{wu2019JointActive}
------, ``Joint active and passive beamforming optimization for intelligent
  reflecting surface assisted {SWIPT} under {QoS} constraints,'' \emph{IEEE J.
  Sel. Areas Commun.}, vol.~38, no.~8, pp. 1735--1748, Aug. 2020.

\bibitem{guan2020intelligent}
X.~{Guan}, Q.~{Wu}, and R.~{Zhang}, ``Intelligent reflecting surface assisted
  secrecy communication: Is artificial noise helpful or not?'' \emph{IEEE
  Wireless Commun. Lett.}, Jan. 2020, {DOI}: 10.1109/LWC.2020.2969629.

\bibitem{huang2020ReconfigurableIntelligent}
C.~{Huang}, R.~{Mo}, and C.~{Yuen}, ``Reconfigurable intelligent surface
  assisted multiuser {MISO} systems exploiting deep reinforcement learning,''
  \emph{IEEE J. Sel. Areas Commun.}, 2020, early access, doi:
  10.1109/JSAC.2020.3000835.

\bibitem{irmer2011coordinated}
R.~Irmer, H.~Droste, P.~Marsch, M.~Grieger, G.~Fettweis, S.~Brueck, H.-P.
  Mayer, L.~Thiele, and V.~Jungnickel, ``Coordinated multipoint: Concepts,
  performance, and field trial results,'' \emph{IEEE Commun. Mag.}, vol.~49,
  no.~2, pp. 102--111, Feb. 2011.

\bibitem{access2010further}
{3GPP TR 36.814}, ``Further advancements for {E-UTRA} physical layer aspects,''
  \emph{{Release 9, v. 9.0.0, Mar}}, 2010.

\bibitem{Ngo2017cell}
H.~Q. {Ngo}, A.~{Ashikhmin}, H.~{Yang}, E.~G. {Larsson}, and T.~L. {Marzetta},
  ``Cell-free massive {MIMO} versus small cells,'' \emph{IEEE Trans. Wireless
  Commun.}, vol.~16, no.~3, pp. 1834--1850, Mar. 2017.

\bibitem{Nayebi2017precoding}
E.~{Nayebi}, A.~{Ashikhmin}, T.~L. {Marzetta}, H.~{Yang}, and B.~D. {Rao},
  ``Precoding and power optimization in cell-free massive {MIMO} systems,''
  \emph{IEEE Trans. Wireless Commun.}, vol.~16, no.~7, pp. 4445--4459, Jul.
  2017.

\bibitem{nigam2014coordinated}
G.~Nigam, P.~Minero, and M.~Haenggi, ``Coordinated multipoint joint
  transmission in heterogeneous networks,'' \emph{IEEE Trans. Commun.},
  vol.~62, no.~11, pp. 4134--4146, Nov. 2014.

\bibitem{tang2018energy}
J.~Tang, A.~Shojaeifard, D.~K. So, K.-K. Wong, and N.~Zhao, ``Energy efficiency
  optimization for {CoMP-SWIPT} heterogeneous networks,'' \emph{IEEE Trans.
  Commun.}, vol.~66, no.~12, pp. 6368--6383, Dec. 2018.

\bibitem{2020Intelligent}
\BIBentryALTinterwordspacing
Q.~Wu, S.~Zhang, B.~Zheng, C.~You, and R.~Zhang, ``Intelligent reflecting
  surface aided wireless communications: A tutorial,'' 2020. [Online].
  Available: \url{https://arxiv.org/abs/2007.02759v1}
\BIBentrySTDinterwordspacing

\bibitem{kang2014joint}
J.~{Kang}, O.~{Simeone}, J.~{Kang}, and S.~S. {Shitz}, ``Joint signal and
  channel state information compression for the backhaul of uplink network
  {MIMO} systems,'' \emph{IEEE Trans. Wireless Commun.}, vol.~13, no.~3, pp.
  1555--1567, 2014.

\bibitem{masoumi2020Performance}
H.~{Masoumi} and M.~J. {Emadi}, ``Performance analysis of cell-free massive
  mimo system with limited fronthaul capacity and hardware impairments,''
  \emph{IEEE Trans. Wireless Commun.}, vol.~19, no.~2, pp. 1038--1053, 2020.

\bibitem{shi2011iteratively}
Q.~Shi, M.~Razaviyayn, Z.-Q. Luo, and C.~He, ``An iteratively weighted {MMSE}
  approach to distributed sum-utility maximization for a {MIMO} interfering
  broadcast channel,'' \emph{IEEE Trans. Signal Process.}, vol.~59, no.~9, pp.
  4331--4340, Sep. 2011.

\bibitem{cvx}
M.~Grant and S.~Boyd, ``{CVX}: Matlab software for disciplined convex
  programming, version 2.1,'' \url{http://cvxr.com/cvx}, Mar. 2014.

\bibitem{boyd2004convex}
S.~Boyd and L.~Vandenberghe, \emph{Convex optimization}.\hskip 1em plus 0.5em
  minus 0.4em\relax Cambridge university press, 2004.

\bibitem{zhang2017matrix}
X.-D. Zhang, \emph{Matrix analysis and applications}.\hskip 1em plus 0.5em
  minus 0.4em\relax Cambridge University Press, 2017.

\bibitem{sun2016majorization}
Y.~Sun, P.~Babu, and D.~P. Palomar, ``Majorization-minimization algorithms in
  signal processing, communications, and machine learning,'' \emph{IEEE Trans.
  Signal Process.}, vol.~65, no.~3, pp. 794--816, Feb. 2016.

\bibitem{song2016Sequence}
J.~{Song}, P.~{Babu}, and D.~P. {Palomar}, ``Sequence design to minimize the
  weighted integrated and peak sidelobe levels,'' \emph{IEEE Trans. Signal
  Process.}, vol.~64, no.~8, pp. 2051--2064, Apr. 2016.

\bibitem{hua2019joint}
M.~Hua, Y.~Wang, M.~Lin, C.~Li, Y.~Huang, and L.~Yang, ``Joint {CoMP}
  transmission for {UAV}-aided cognitive satellite terrestrial networks,''
  \emph{IEEE Access}, vol.~7, pp. 14\,959--14\,968, 2019.

\bibitem{wang2015energy}
Y.~Wang, C.~Li, Y.~Huang, D.~Wang, T.~Ban, and L.~Yang, ``Energy-efficient
  optimization for downlink massive {MIMO} {FDD} systems with transmit-side
  channel correlation,'' \emph{{IEEE} Trans. Veh. Technol.}, vol.~65, no.~9,
  pp. 7228--7243, Sep. 2015.

\bibitem{lobo1998applications}
M.~S. Lobo, L.~Vandenberghe, S.~Boyd, and H.~Lebret, ``Applications of
  second-order cone programming,'' \emph{Linear algebra and its applications},
  vol. 284, no. 1-3, pp. 193--228, Nov. 1998.

\bibitem{sidiropoulos2006transmit}
N.~D. Sidiropoulos, T.~N. Davidson, and Z.-Q. Luo, ``Transmit beamforming for
  physical-layer multicasting,'' \emph{IEEE Trans. Signal Process.}, vol.~54,
  no.~6, pp. 2239--2251, Jun. 2006.

\bibitem{pan2017joint}
C.~Pan, H.~Zhu, N.~J. Gomes, and J.~Wang, ``Joint precoding and {RRH} selection
  for user-centric green {MIMO} {C-RAN},'' \emph{IEEE Trans. Wireless Commun.},
  vol.~16, no.~5, pp. 2891--2906, May 2017.

\bibitem{ebj2020Intelligent}
E.~{Bj{\"o}rnson}, {\"O}.~{{\"O}zdogan}, and E.~G. {Larsson}, ``Intelligent
  reflecting surface versus decode-and-forward: How large surfaces are needed
  to beat relaying?'' \emph{IEEE Wireless Commun. Lett.}, vol.~9, no.~2, pp.
  244--248, 2020.

\bibitem{direnzo2020Reconfigurable}
M.~{Di Renzo}, K.~{Ntontin}, J.~{Song}, F.~H. {Danufane}, X.~{Qian},
  F.~{Lazarakis}, J.~{De Rosny}, D.~{Phan-Huy}, O.~{Simeone}, R.~{Zhang},
  M.~{Debbah}, G.~{Lerosey}, M.~{Fink}, S.~{Tretyakov}, and S.~{Shamai},
  ``Reconfigurable intelligent surfaces vs. relaying: Differences,
  similarities, and performance comparison,'' \emph{IEEE Open Journal of the
  Communications Society}, vol.~1, pp. 798--807, 2020.

\end{thebibliography}
%\begin{thebibliography}{1}	
\begin{IEEEbiography}[{\includegraphics[width=1in,height=1.25in,clip,keepaspectratio]{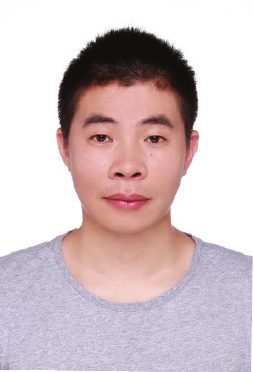}}]{Meng Hua}
	received the  M.S. degree in electrical and information engineering from Nanjing University of Science and Technology, Nanjing, China, in 2016. Since September 2016, he is currently working towards the Ph.D. degree in School of Information Science and Engineering, Southeast University, Nanjing, China. His current research interests include UAV assisted communication, intelligent reflecting surface (IRS), backscatter communication,  energy-efficient wireless communication, X-connectivity, cognitive radio network, secure transmission, and optimization theory.
\end{IEEEbiography}

\begin{IEEEbiography}[{\includegraphics[width=1in,height=1.25in,clip,keepaspectratio]{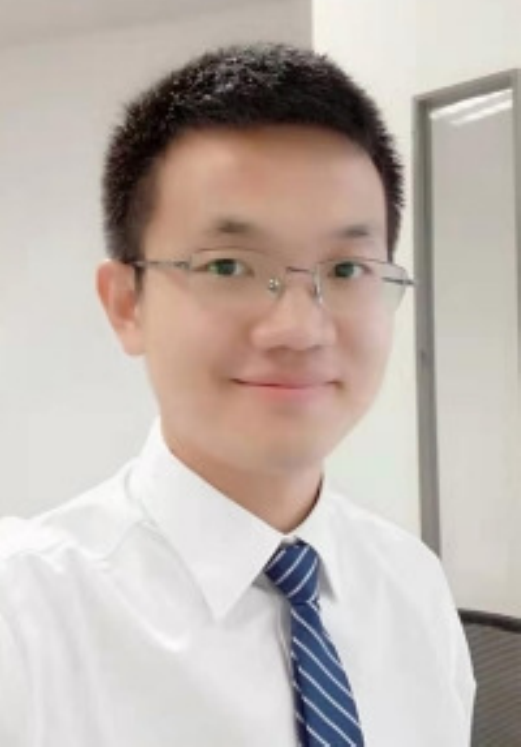}}] {Qingqing Wu} (S'13-M'16) received the B.Eng. and the Ph.D. degrees in Electronic Engineering from South China University of Technology and Shanghai Jiao Tong University (SJTU) in 2012 and 2016, respectively. He is currently an Assistant Professor in the Department of Electrical and Computer Engineering at the University of Macau, China, and also with the State key laboratory of Internet of Things for Smart City. He was a Research Fellow in the Department of Electrical and Computer Engineering at National University of Singapore. His current research interest includes intelligent reflecting surface (IRS), unmanned aerial vehicle (UAV) communications, and MIMO transceiver design. He has published over 80 IEEE journal and conference papers.
	
	He was the recipient of the IEEE WCSP Best Paper Award in 2015, the Outstanding Ph.D. Thesis Funding in SJTU in 2016, the Outstanding Ph.D. Thesis Award of China Institute of Communications in 2017. He was the Exemplary Editor of IEEE Communications Letters in 2019 and the Exemplary Reviewer of several IEEE journals. He serves as an Associate Editor for IEEE Communications Letters and IEEE Open Journal of Communications Society. He is the Lead Guest Editor for IEEE Journal on Selected Areas in Communications on ``UAV Communications in 5G and Beyond Networks", and the Guest Editor for IEEE Open Journal on Vehicular Technology on ``6G Intelligent Communications" and IEEE Open Journal of Communications Society on ``Reconfigurable Intelligent Surface-Based  Communications for 6G Wireless Networks". He is the workshop co-chair for ICC 2019 and ICC 2020 workshop on ``Integrating UAVs into 5G and Beyond", and the workshop co-chair for GLOBECOM 2020 workshop on ``Reconfigurable Intelligent Surfaces for Wireless Communication for Beyond 5G". 
\end{IEEEbiography}

\begin{IEEEbiography}[{\includegraphics[width=0.95in,height=1.4in]{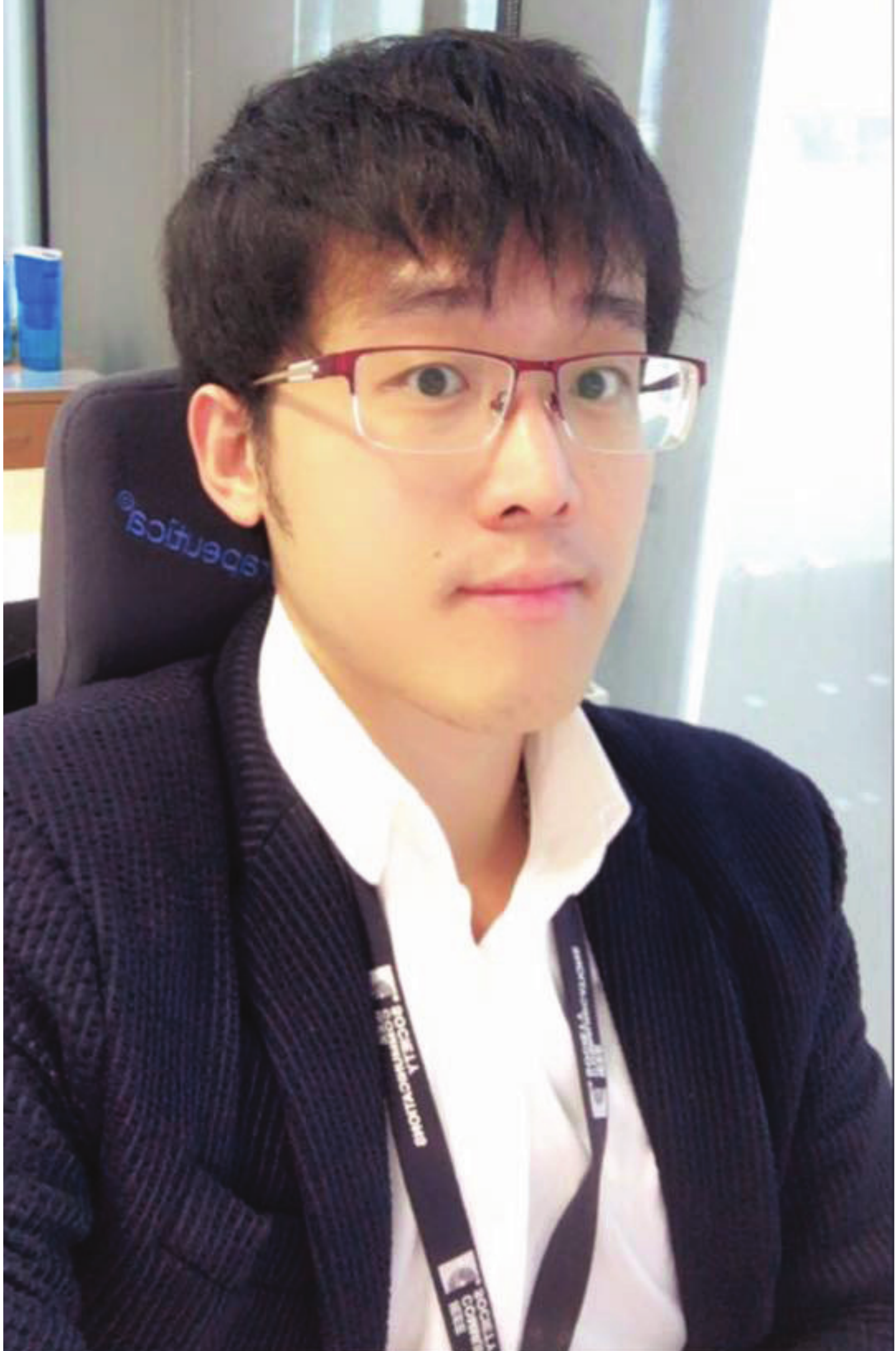}}]{Derrick
		Wing Kwan Ng   }(S'06-M'12-SM'17-F'21) received the bachelor degree with first-class honors and the Master of Philosophy (M.Phil.) degree in electronic engineering from the Hong Kong University of Science and Technology (HKUST) in 2006 and 2008, respectively. He received his Ph.D. degree from the University of British Columbia (UBC) in 2012. He was a senior postdoctoral fellow at the Institute for Digital Communications, Friedrich-Alexander-University Erlangen-N\"urnberg (FAU), Germany. He is now working as a Senior Lecturer and a Scientia Fellow at the University of New South Wales, Sydney, Australia.  His research interests include convex and non-convex optimization, physical layer security, IRS-assisted communication, UAV-assisted communication, wireless information and power transfer, and green (energy-efficient) wireless communications. 
	
	Dr. Ng received the Australian Research Council (ARC) Discovery Early Career Researcher Award 2017,   the Best Paper Awards at the IEEE TCGCC Best Journal Paper Award 2018, INISCOM 2018, IEEE International Conference on Communications (ICC) 2018,  IEEE International Conference on Computing, Networking and Communications (ICNC) 2016,  IEEE Wireless Communications and Networking Conference (WCNC) 2012, the IEEE Global Telecommunication Conference (Globecom) 2011, and the IEEE Third International Conference on Communications and Networking in China 2008.  He has been serving as an editorial assistant to the Editor-in-Chief of the IEEE Transactions on Communications from Jan. 2012 to Dec. 2019. He is now serving as an editor for the IEEE Transactions on Communications,  the IEEE Transactions on Wireless Communications, and an area editor for the IEEE Open Journal of the Communications Society. Also, he has been listed as a Highly Cited Researcher by Clarivate Analytics since 2018. He is an IEEE Fellow.
\end{IEEEbiography}

\begin{IEEEbiography}[{\includegraphics[width=1in,height=1.25in,clip,keepaspectratio]{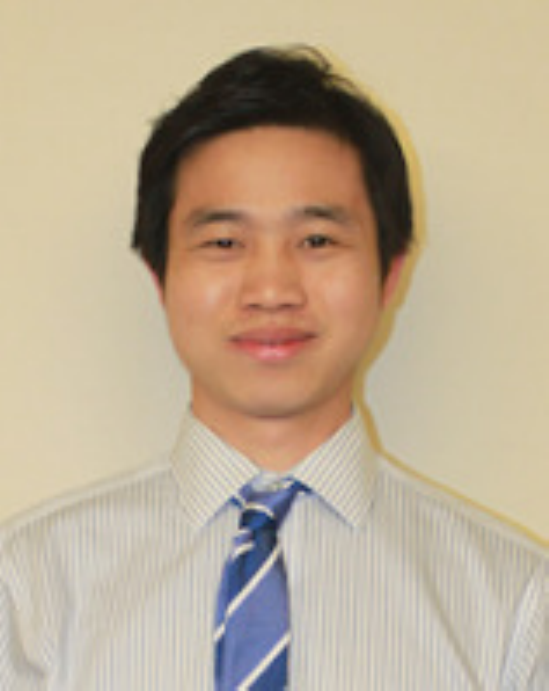}}]{Jun Zhao}
	 (S'10-M'15) is currently an Assistant Professor in the School of Computer Science and Engineering at Nanyang Technological University (NTU) in Singapore. He received a PhD degree in May 2015 in Electrical and Computer Engineering from Carnegie Mellon University (CMU) in the USA (advisors: Virgil Gligor, Osman Yagan; collaborator: Adrian Perrig), affiliating with CMU's renowned CyLab Security \& Privacy Institute, and a bachelor's degree in July 2010 from Shanghai Jiao Tong University in China. Before joining NTU first as a postdoc with Xiaokui Xiao and then as a faculty member, he was a postdoc at Arizona State University as an Arizona Computing PostDoc Best Practices Fellow (advisors: Junshan Zhang, Vincent Poor). His research interests include communication networks, security/privacy, and AI. His co-authored papers received Best Paper Award (IEEE Transaction Paper) by IEEE Vehicular Society (VTS) Singapore Chapter in 2019, and Best Paper Award in EAI International Conference on 6G for Future Wireless Networks (EAI 6GN) 2020.
\end{IEEEbiography}

\begin{IEEEbiography}[{\includegraphics[width=1in,height=1.25in,clip,keepaspectratio]{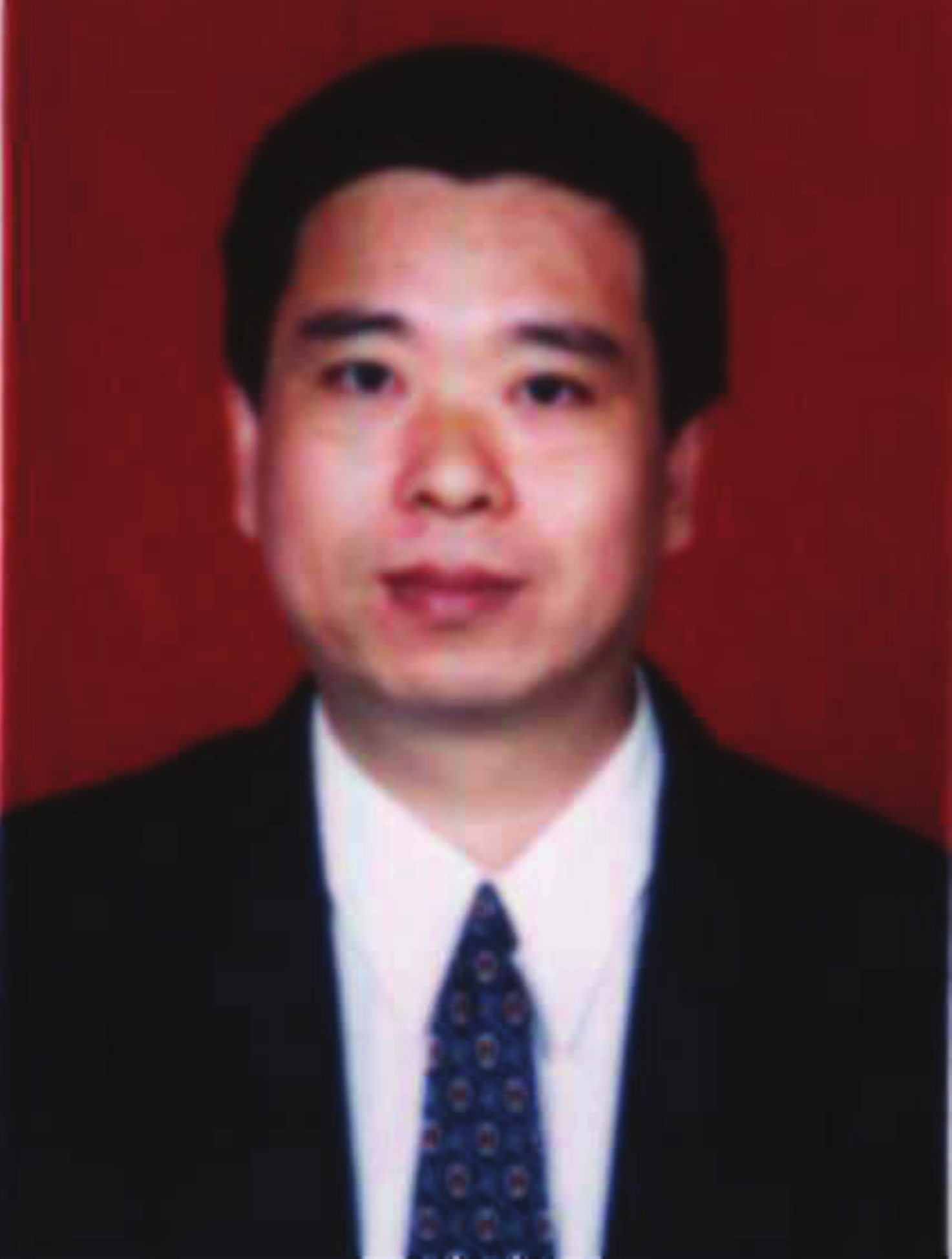}}]{Luxi Yang}
	(M'96-SM'17) received the M.S. and Ph.D. degrees in electrical engineering from Southeast University, Nanjing, China, in 1990 and 1993, respectively. Since 1993, he has been with the Department of Radio Engineering, Southeast University, where he is currently a Full Professor of information systems and communications, and the Director of the Digital Signal Processing Division. He has authored or co-authored of two published books and more than 200 journal papers, and holds 50 patents. His current research interests include signal processing for wireless communications, MIMO communications, intelligent wireless communications, and statistical signal processing. He received the first and second class prizes of science and technology progress awards of the State Education Ministry of China in 1998, 2002, and 2014. He is currently a member of Signal Processing Committee of the Chinese Institute of Electronics.
\end{IEEEbiography}
\end{document}